\documentclass[12pt,twocol]{aa}

\usepackage{hyperref}
\usepackage{graphicx,amssymb,amsmath,times}
\graphicspath{{../../../Surveys/VVV/data/}}
\usepackage{longtable}
\usepackage{bm} 
\usepackage{url}
\usepackage[varg]{txfonts}
\usepackage[T1]{fontenc}
\usepackage{natbib}
\bibpunct{(}{)}{;}{a}{}{,} 

\usepackage[normalem]{ulem}

\def\simlt{\lower.5ex\hbox{\ltsima}}
\def\simgt{\lower.5ex\hbox{\gtsima}}

\def\gtsim{\;\lower.6ex\hbox{$\sim$}\kern-6.7pt\raise.4ex\hbox{$>$}\;}
\def\ltsim{\;\lower.6ex\hbox{$\sim$}\kern-6.9pt\raise.4ex\hbox{$<$}\;}

\def\Deg{${}^\circ$\llap{.}}

\def\Sec{${}^{\prime\prime}$\llap{.}}
\def\deg{${}^\circ$}
\def\min{${}^{\prime}$}
\def\sec{${}^{\prime\prime}$}

\def\jmk{\hbox{\it J--K$_s$\/}}

\titlerunning{Type II Cepheids in the Galactic bulge based on near-infrared data}
\authorrunning{Braga et al.}

\begin{document}

\title{Structure and kinematics of Type II Cepheids in the Galactic bulge based on near-infrared VVV data}


   \author{V.~F.~Braga\inst{1,2} \and A.~Bhardwaj\inst{3} \and R.~Contreras Ramos\inst{1,4}
   \and D.~Minniti\inst{1,2,5} \and G.~Bono\inst{6,7} \and R.~de Grijs\inst{8,9} \and J.~H.~Minniti\inst{4,10} \and M.~Rejkuba\inst{11}}

   \institute{Instituto Milenio de Astrof{\'i}sica, Santiago, Chile\\
         \and
Departamento de F{\'i}sica, Facultad de Ciencias Exactas, Universidad Andr{\'e}s Bello, Fern{\'a}ndez Concha 700, Las Condes, Santiago, Chile\\
         \and
Kavli Institute for Astronomy and Astrophysics, Peking University, Yi He Yuan Lu 5, Hai Dian District, Beijing 100871, China\\
         \and
Pontificia Universidad Cat{\'o}lica de Chile, Instituto de Astrof{\'i}sica, Av. Vicu\~na Mackenna 4860, Macul, Chile\\
         \and
Vatican Observatory, V00120 Vatican City State, Italy\\
         \and
Department of Physics, Universit\`a di Roma Tor Vergata, via della Ricerca Scientifica 1, 00133 Roma, Italy\\
         \and
INAF-Osservatorio Astronomico di Roma, via Frascati 33, 00040 Monte Porzio Catone, Italy\\
         \and
Department of Physics and Astronomy, Macquarie University, Balaclava Road, Sydney, NSW 2109, Australia\\
         \and
International Space Science Institute--Beijing, 1 Nanertiao, Zhongguancun, Hai Dian District, Beijing 100190, China\\
         \and
European Southern Observatory, Alonso de C{\'o}rdova 3107, Casilla 19001, Santiago, Chile\\
         \and
European Southern Observatory, Karl-Schwarzschild-Stra\ss e 2, 85748, Garching, Germany}

\date{\centering Submitted \today\ / Received / Accepted }

\abstract
{Type II Cepheids (T2Cs) are radially pulsating variables that trace old stellar populations and 
provide distance estimates through their period-luminosity (PL) relation.}
{We trace the structure of old stellar population in the Galactic bulge 
using new distance estimates and kinematic properties of T2Cs.}
{We present new near-infrared photometry of T2Cs in the bulge from 
the VISTA Variables in the V{\'i}a L{\'a}ctea survey (VVV). We provide the largest
sample (894 stars) of T2Cs with $JHK_s$ observations that have 
accurate periods from the Optical Gravitational Lensing Experiment (OGLE) catalog. 
Our analysis makes use of the $K_s$-band time-series observations to estimate mean magnitudes and
individual distances by means of the PL relation. To constrain the
kinematic properties of our targets, we complement our analysis
with proper motions based on both the VVV and {\it Gaia} Data Release 2.}
{We derive an empirical $K_s$-band PL relation that depends on Galactic longitude and latitude: 
$K_{s0} = (10.66\pm0.02) - (2.21\pm0.03)\cdot(\log{P}-1.2) - (0.020\pm0.003)\cdot l + (0.050\pm0.008)\cdot |b|$
mag; individual extinction corrections 
are based on a 3D reddening map. Our targets display a 
centrally concentrated distribution, with solid evidence of ellipsoidal
symmetry---similar to the RR Lyr{\ae} ellipsoid---and a few halo outliers
up to $\gtrsim$100 kpc. 
We obtain a distance from the Galactic center of $R_0$=8.46$\pm$0.03(stat.)$\pm$0.11(syst.) kpc. 
We also find evidence that the bulge T2Cs belong to a kinematically hot population,  
as the tangential velocity components ($\sigma v_{l*}$=104.2$\pm$3.0 
km/s and $\sigma v_{b}$=96.8$\pm$5.5 km/s) agree 
within 1.2$\sigma$. Moreover, the difference between absolute and relative proper motion is in good
agreement with the proper motion of Sgr A* from VLBA measures.}
{We conclude that bulge T2Cs display an ellipsoidal
spatial distribution and have kinematics similar to 
RR Lyr{\ae} stars, which are other tracers
of the old, low-mass stellar population. 
T2Cs also provide an estimate of $R_0$ that agrees excellently well with the literature, taking 
account of the reddening law.}

\maketitle

\section{Introduction}\label{section:intro}


In the Galactic bulge, the Red Clump (RC) stars, which are core-helium burning low-mass stars
with ages from intermediate (1$\leq$age<10 Gyr) to old (age$\geq$10 Gyr) and average to high 
metallicities \citep[{[Fe/H]}$\gtsim$--1.5 dex,][]{cole1998,hill2011}, are 
used extensively to study their kinematics, chemical abundances, and spatial 
distributions \citep{mcwilliamzoccali2010,saito2011,gonzalez2015, zoccali2017}. 
High-resolution spectroscopic studies of RCs suggest that only stars
that are more metal-rich than [Fe/H]$\sim$--0.5 dex trace an X-shaped structure
that appears to be in the form of a peanut-boxy bulge 
\citep{ness2013,zoccali2016}. Recently, mid-infrared 
data obtained with the Wide-field Infrared Survey Explorer (WISE) also showed a clear large-scale X structure \citep{nesslang2016}. 
In contrast, stars that are more metal-poor than [Fe/H]$\sim$--0.5 dex not only display a 
centrally concentrated axisymmetric spatial distribution, but also reveal 
different kinematics \citep{ness2013,valenti2016,zoccali2017}.

However, the spatial distribution of the old and more metal-poor population of stars in the bulge, 
which is traced by RR Lyr{\ae} (RRLs), is still under discussion. \citet{pietrukowicz2015} found an
ellipsoidal distribution (axis ratios of 1 : 0.49$\pm$0.02 : 0.39$\pm$0.02), elongated 
along the same direction as the bar traced by the metal-rich red giants, 
while \citet{dekany2013} and \citet{kunder2016} found a spheroidal distribution.

Type II Cepheids (T2Cs) are old (>10 Gyr), low-mass
post-horizontal branch, asymptotic giant branch (AGB) and post-AGB stars. 
Like the RRLs, T2Cs trace old stellar 
populations, but they have longer periods (1-80 days), are brighter by 
1-3 mag, and their amplitudes can be up to twice as large as those of the RRLs.
In his pioneering work that led to the separation
of Population I and Population II stars, \citet{baade44} showed that
T2Cs are distance indicators that obey a period-luminosity (PL) relation
different from that of Classical Cepheids. 
T2Cs have been widely used in the literature as distance indicators, 
both in the optical \citep{harris85,nemec1994} and in the near-infrared (NIR) 
bands \citep{matsunaga06,matsunaga11a,bhardwaj17b,bhardwaj17c}, although 
not as frequently as RRLs and Classical Cepheids.
A key feature of the PL relations is that their intrinsic dispersion becomes 
smaller from the optical to the NIR
\citep{dicriscienzo07}. This means that NIR PL relations are not only more
accurate because the reddening is less severe, but they are also intrinsically
more precise.

Near-infrared photometry of T2Cs in the bulge obtained with the Son of ISAAC (SOFI) telescope was used by
\citet{groenewegen08} to estimate the distance of the Galactic center 
($R_0$=7.99$\pm$0.09 kpc) using a sample of 38 T2Cs. 
\citet{bhardwaj17c} matched photometry of the VISTA Variables 
in the V\'ia L\'actea (VVV) survey 
\citep{minniti2010,saito2012} with the 
OGLE III version of the catalog of T2Cs \citep[][335 T2Cs]{soszynski2011} and obtained
individual distances. They estimated $R_0$=8.34$\pm$0.03 kpc 
and ruled out a barred structure. All recent estimates of $R_0$ based on 
other diagnostics and recent reviews agree that the official IAU value of $R_0$=8.5 kpc
is overestimated ($R_0$=8.2$\pm$0.1 kpc, \citealt{blandhawthorn2016};
$R_0$=8.3$\pm$0.2$\pm$0.4 kpc, \citealt{degrijs2016}).

Recently, the Optical Gravitational Lensing Experiment (OGLE) IV survey \citep{udalski2015} 
has generated the largest homogeneous sample
of T2Cs known to date, amounting to 924 objects projected toward the Galactic bulge
\citep{soszynski2017}. This is almost three times the size of the previous sample.
The VVV survey has collected NIR $K_s$-band time series toward the Galactic bulge in a sky 
area that covers almost the entire OGLE survey area 
and provides an optimal framework to characterize 
the structure of the old population of 
the Galactic bulge with stellar tracers such as RRLs and T2Cs, 
for which the optical photometry and accurate periods are available from the OGLE survey. 
The increase of the sample size with respect to previous works 
is a unique opportunity
to achieve new insight into the old stellar population in the bulge, 
especially for a detailed comparison with RRLs, which has
always been hampered by the small sample size of T2Cs.
Furthermore, we have the unprecedented opportunity to 
combine the T2C NIR catalog 
with the proper motion measurements from VVV itself \citep{contreras2017,smith2018}
and {\it Gaia} DR2 \citep{gaia_alldr,gaia_dr2} to constrain the
kinematic properties of the old stellar population.

The paper is organized as follows: in Section~\ref{section:data} we present our photometric and astrometric databases. We analyze the
light curves and derive their properties in Section~\ref{section:lcv}.
Section~\ref{section:pls} is dedicated to estimating individual 
distances of T2Cs and their overall distribution, while in Section~\ref{section:kinematics}
we discuss the kinematic properties of our targets. We 
discuss and summarize our results in Section~\ref{section:conclusions}.

\section{Data}\label{section:data}

{\it Light curves.}  We used the aperture photometry of VVV DR4 
data \citep{minniti2010,saito2012} that is publicly 
available through VISTA Science Archive (VSA)\footnote{\url{vsa.roe.ac.uk/index.html}}. 
We compared point spread function (PSF) with aperture photometry for 
a sample of our targets and found the differences to be  negligible, but 
aperture photometry has the advantage of being available in
the entire VVV survey area.
As a first step, we matched the OGLE IV catalog of T2Cs 
\citep{soszynski2017} with the source detection catalog
of VVV, adopting a matching radius of 2{\sec}. This allowed us to 
retrieve 894 of 924 targets within the VVV survey area. A posteriori,
we checked that all the good matches are within 1\Sec3 of the 
OGLE coordinates. Of those that were not retrieved, 25 are outside 
the VVV area, and for five of them we could not find a 
good match, even with a larger searching radius of 10{\sec}.
Of these 894 T2Cs, according to the classification of \citet{soszynski2017}, 
369 are BL Herculis (BLHs), 343 are W Virginis (WVs), 28 are peculiar
W Virginis (pWVs), and 154 are RV Tauri (RVTs). We  discuss  
the different types of T2Cs in more detail in Section~\ref{section:dist}.

We collected both the single-epoch $JH$-band photometry and 
the $K_s$-band time series. $ZY$ photometry
was neglected because it is not useful for our goals and 
we cannot even estimate mean magnitudes in these bands, since
light-curve templates \citep{bhardwaj17b} are available only for the $JHK_s$ bands.
The number of valid $K_s$-band phase points per variable (those with
good photometric solution) ranges from four 
to 185, with a median of 51. Only four variables have 
fewer than ten phase points, and 90\% of the variables have more than 
47 observations, with a good phase coverage over the whole range of 
periods of our targets (1-85 days). The number of available $K_s$ epochs 
varies across the VVV survey area, with the majority of the VVV pointings 
(so-called tiles) having between 50-100 epochs. Given the overlap 
between the adjacent tiles, which amounts to about 1{\min} on a side, a small 
fraction of our variables (34) were observed in two tiles, with up to 185 data points.

{\it Reddening.} We adopted the two Galactic bulge E(\jmk) reddening maps of
\citet{gonzalez2012} and \citet{schultheis2014}, henceforth, G12 and S14. The reasons are
manifold: {\it 1)} They were both obtained with VVV data, therefore
no photometric system conversion is needed. {\it 2)} The map of 
G12 provides very high resolution especially in the 
central regions (2$\times$2 arcmin for --3\Deg5<$b$<5\Deg0, 4$\times$4 arcmin for --7\Deg0<$b$<--3\Deg5
and 6$\times$6 arcmin for --10\Deg0<$b$<--7\Deg0), while the ``pixels'' of the 
S14 map have a size of 6$\times$6 arcmin everywhere. However, the map by
S14 has the key advantage to be three-dimensional.
It provides a grid of E(\jmk) for 21 bins of distance, from 
0 to 10.5 kpc. This is crucial for studying a complex structure such
as that of the Galactic bulge and possibly intervening thick-disk populations. 
In the following, we indicate as 
E(\jmk)$_{G12}$ and E(\jmk)$_{S14}$ the two reddening values
obtained from the G12 and S14 maps, respectively. As explained 
in Section~\ref{section:lcv}, we adopt the 3D reddening map of S14
for our final estimates, but G12 serves as a comparison.

{\it Proper motions.}  We retrieved relative proper 
motions for 894 of 924 targets from 
the publicly available VIRAC catalog \citep{smith2018}, obtained with VVV data. According to their recommendations, 
we discarded all targets for which the flag $reliable$, based on 
the validation of the photometric solutions, is equal to zero.
We point out that although the cross-match was made by unique VVV ID and 
not by coordinate, we found multiple 
(either double, triple, or quadruple) records for 120 targets in VIRAC. We checked
the multiple identifications one by one on the basis of right ascension ($\alpha$),
declination ($\delta$), and 
$K_s$-band magnitude. About two-thirds of the time, all of the records of 
a multiple identification had $reliable$=0 and were thus all discarded. Of the remaining fraction, 
an a posteriori check revealed that the majority had $reliable$=1 for the correct 
match and $reliable$=0 for the incorrect matches. This validates our
selection of the best match.

We also retrieved relative proper motions
for 416 targets from PSF photometry, obtained from
VVV data by the method explained in \citet{contreras2017}. 
The match was performed using a searching radius of 2\sec.
This catalog of proper motions does not cover the entire VVV
area, but only the low latitudes (--3\Deg0$\ltsim b \ltsim$3\Deg0), that is, 
the most crowded region, where PSF photometry has several advantages
over aperture photometry. A comparison between the two sets of
proper motions from the VVV is performed in Section~\ref{section:kinematics}.

Finally, we searched our targets within the recent {\it Gaia} DR2 
\citep{gaia_alldr,gaia_dr2}, using a 
searching radius of 4\sec. We  retrieved matches for
920 targets. Because of the density of {\it Gaia} DR2,
a search radius of 4{\sec} means multiple records for almost 
all our targets. We note that for {\it Gaia}, we needed a larger searching 
radius because $\alpha$ and $\delta$ are at the epoch J2015.5. 
We selected the best matches on the basis of
the separation in ($\alpha$,$\delta$) and the $V-G$ and $I-G$ color indexes, 
and retrieved 914 targets (for the remaining 6 targets, it was not possible to select
a best match based on the adopted criteria). 
Of these, 868 have a five-parameter {\it Gaia} solution (coordinates, 
parallaxes, and absolute proper motions). 


\section{Light curves and properties of T2Cs}\label{section:lcv}

We have phased the VVV light curves using the periods 
provided by OGLE IV \citep{soszynski2017}. We visually 
inspected all the VVV light curves and separated promising from noisy or poorly sampled light curves.
We estimated the uncertainty on the mean magnitude ($eK_s$)
as the sum in quadrature of the standard error of mean of the phase 
points around the fit (see below) plus the median photometric 
error of the phase points. We note that the standard error 
on the mean and median photometric error represents a statistical 
and systematic measure of uncertainties, respectively,

$$eK_s = \sqrt{\dfrac{1}{n}\dfrac{\sum_{i=1}^{n} (mag_i-mag_{i(fit)})^2}{n-1}+median(err)^2.}$$

\begin{figure*}[!htbp]
\centering
\includegraphics[height=8cm]{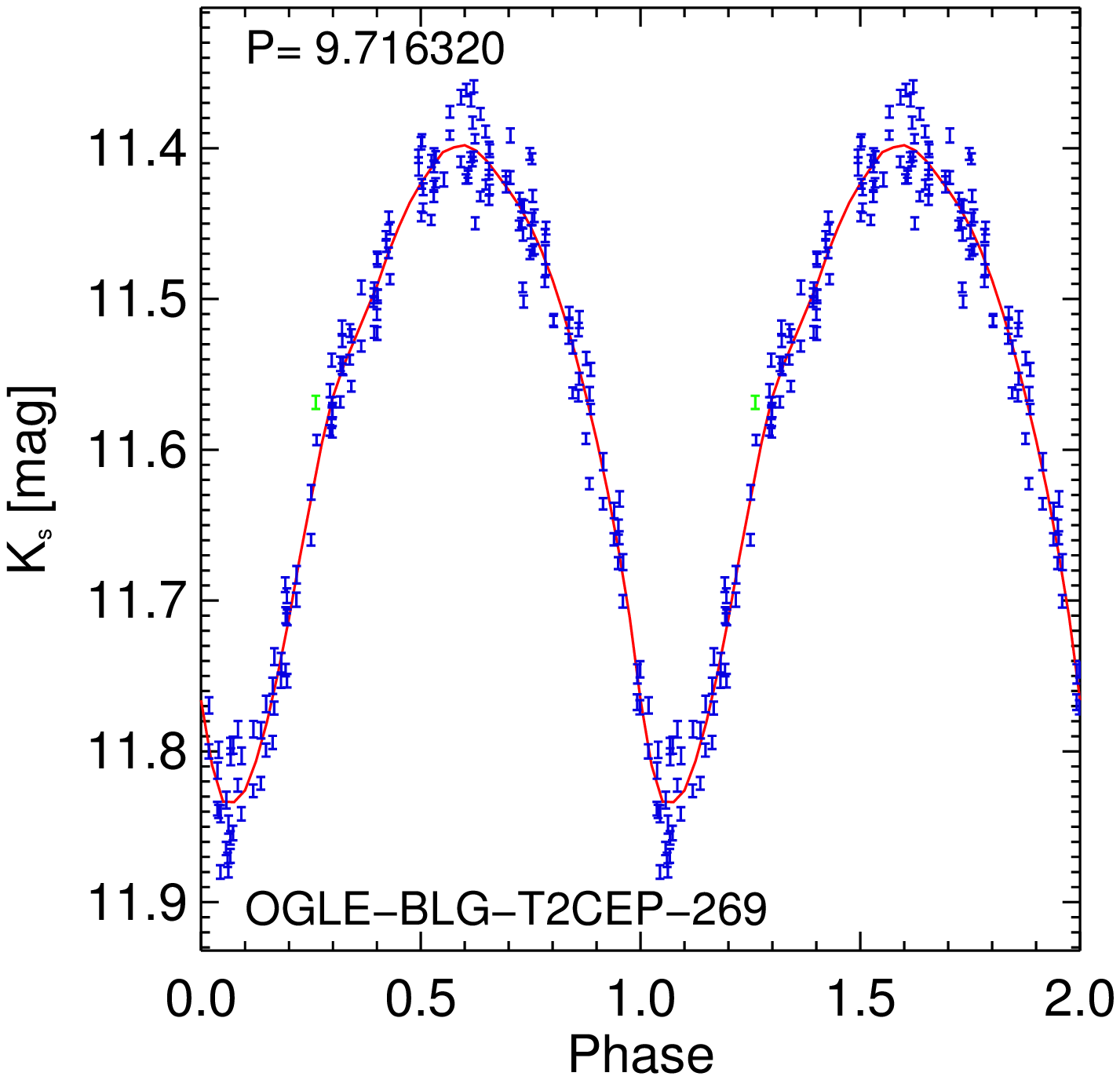}
\includegraphics[height=8cm]{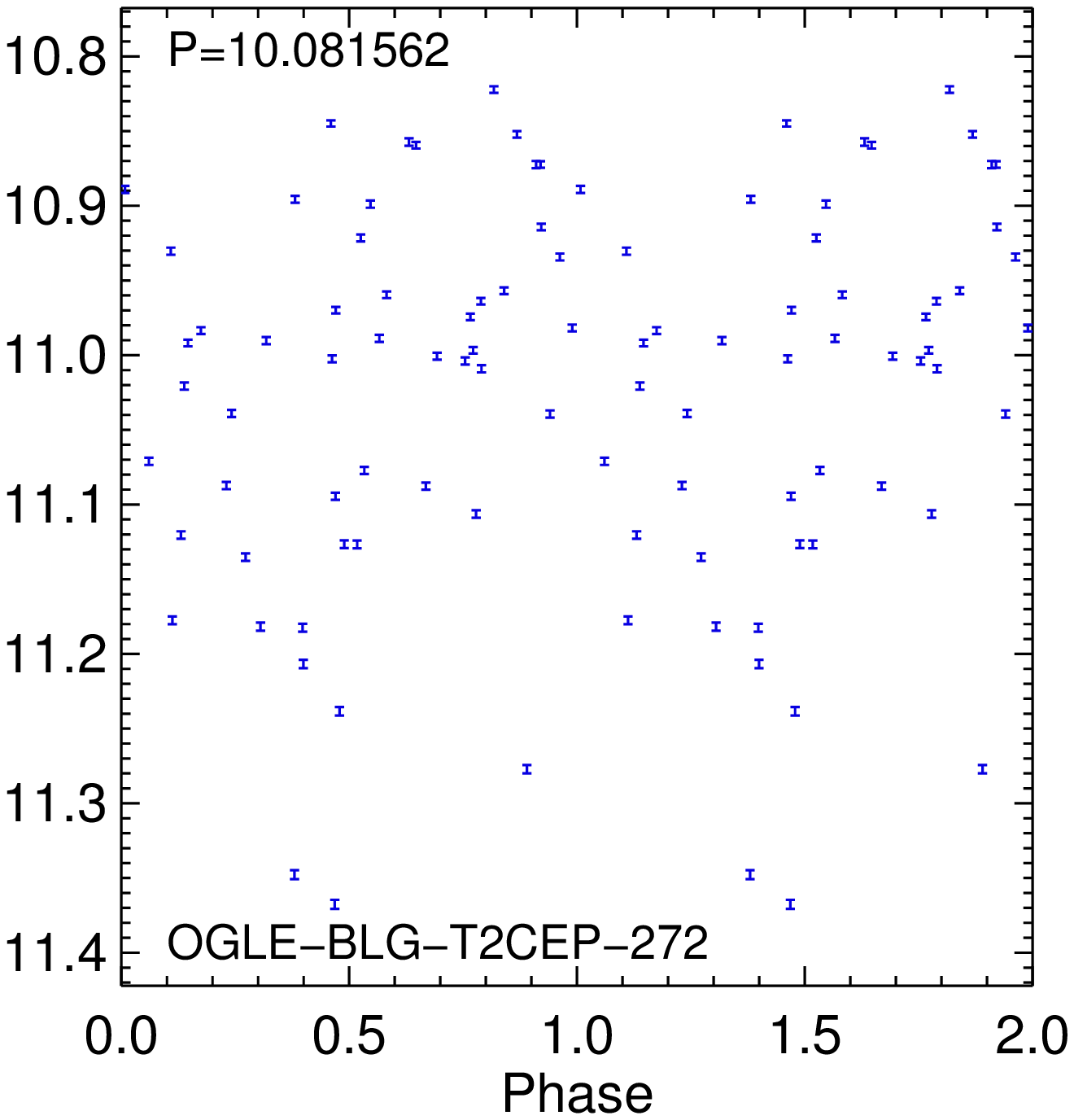}
\caption{Left: Example of a good-quality light curve.
Right: Example of a poor-quality light curve. The periods are labeled at
the top of the panels, while the names, as provided in the OGLE 
catalog, are labeled at the bottom. The PLOESS fit of the 
good-quality light curve is displayed as a red line.}
\label{fig:lc_eg}
\end{figure*}

We adopted
$eK_s$ as a quantitative criterion to select good-quality
light curves, with a threshold at $eK_s<0.05$. 
We performed visual inspection to validate this threshold 
and further selected promising light curves within the
range $0.05<eK_s <0.10$~mag. However, some of the brighter 
stars ($\langle K_s\rangle \lesssim 11.5$~mag) were classified as poor-quality 
light curves in this range (see below), and all stars with
$eK_s >0.10$ mag were also included in the poor-quality 
sample. Examples of good- and poor-quality light curves are displayed in Figure~\ref{fig:lc_eg}.

For the 161 targets with poor-quality light curves, we
adopted for the mean magnitude the 
median of the magnitudes, converted into flux, of the individual phase points.
The uncertainty on the mean magnitude is the 
standard deviation of the median. We do not provide 
a light-amplitude estimate for
these targets since the uncertainty would be on the same order
as the amplitude itself. We point out that most of these targets
are very bright (135 of 161 have $\langle K_s\rangle \lesssim 11.5$~mag), meaning that they are either
saturated or in the nonlinear regime of the camera. However, 
we also point out that not all stars in this magnitude range have 
poor-quality light curves. The cause of poor photometry is a synergy 
between saturation and crowding by other nearby bright stars. The magnitude for saturated sources is derived from the most external 
apertures (those with the largest radii). When another bright source contaminates 
the external aperture, the photometric solution is worse. In these cases, PSF photometry
would not help to improve the photometry because it would be even more
affected by saturation.

To compute the mean magnitudes, amplitudes, and relative uncertainties
of the good-quality targets, we adopted
the fits of the light curves obtained with the PLOESS \citep{braga2018} 
fitting method, which is a variant of the GLOESS method \citep{persson2004}.
The mean magnitude is the integral of the fitting curve converted
into flux, while the amplitude is the difference between the brightest and
faintest points of the fit. The uncertainty on the mean magnitude was
defined before as $eK_s$. The uncertainty on the amplitude
was derived as the sum in quadrature of the median photometric errors 
of the phase points around the maximum and minimum, plus the standard deviation of
these phase points around the fit of the light curve. The final
value was weighted with the number of phase points around the 
maximum and the minimum. We also derived mean $J$- and $H$-band
magnitudes by applying the light-curve templates of \citet{bhardwaj17b}
to the single $J$ and $H$ phase points and using the time of maximum
from OGLE \citep{soszynski2017}. The photometric properties of our 
final sample are listed in Table~\ref{tab:jhk}.

\begin{table*}
 \scriptsize
 \caption{Photometric properties of target T2Cs.}
 \centering
 \begin{tabular}{l l l l c c c c c c c}
 \hline\hline  
 ID (OGLE IV)\tablefootmark{a} & ID (VVV) & ID ({\it Gaia}) & Type  & Period & $\langle V \rangle$ & $\langle I \rangle$ & $\langle J \rangle$ & $\langle H \rangle$ & $\langle K_s \rangle$ & Amp($K_s$) \\
  & & & & days & mag & mag & mag & mag & mag & mag \\
\hline
 001 &     \ldots    & 5980064527510861824  & BLHer & 3.9983508  &  15.759 &  14.176  & \ldots  &  \ldots  & \ldots &   \ldots \\
 002 & 515601356315  & 5979966877097299328  & BLHer & 2.2684194  &  15.188 &  13.909  &  12.976$\pm$0.006 &   12.644$\pm$0.007 & 12.507$\pm$0.004 &   0.212$\pm$0.013 \\
 003 & 515601679485  & 5979980380479944704  & BLHer & 1.4844493  &  16.519 &  15.061  &  14.027$\pm$0.007 &   13.648$\pm$0.009 & 13.409$\pm$0.008 &   0.334$\pm$0.019 \\
 004 &      \ldots   & 4107385973738284672  & BLHer & 1.2118999  &  16.404 &  14.856  & \ldots  &  \ldots  & \ldots &   \ldots \\
 005 & 515594023082  & 5980291576574487808  & BLHer & 2.0075505  &  18.666 &  16.842  &  15.485$\pm$0.024 &   14.993$\pm$0.034 & 14.782$\pm$0.034 &   0.407$\pm$0.040 \\
 006 &     \ldots    & 4107420780154292480  & pWVir & 7.6364832  &  14.728 &  13.352  & \ldots  &  \ldots  & \ldots &   \ldots \\
 007 & 515543342733  & 4059723965740384128  & BLHer & 1.8174741  &  17.452 &  15.530  &  13.696$\pm$0.010 &   13.559$\pm$0.012 & 13.292$\pm$0.011  &   0.286$\pm$0.014 \\
 008 & 515520862858  & 4059946032759214976  & BLHer & 1.1829551  &  17.765 &  15.970  & \ldots  &   14.034$\pm$0.015 & 13.967$\pm$0.017 &   0.183$\pm$0.019 \\
 009 & 515555436341  & 4059508427192076928  & BLHer & 1.8960190  &  17.620 &  15.657  &  14.003$\pm$0.009 &   13.457$\pm$0.011 & 13.266$\pm$0.013  &   0.336$\pm$0.029 \\
 010 &    \ldots     & 4109969452496082432  & BLHer & 1.9146565  &  16.639 &  14.969  & \ldots  &  \ldots  & \ldots &   \ldots \\
\hline
 \end{tabular}
\tablefoot{Only the first 10 of the 924 lines of the table are shown. The full table is
 shown in the machine-readable version of the paper.\\
 \tablefoottext{a}{The full name is OGLE-BLG-T2CEP-XXX, where ``XXX''
 is the ID appearing in the first column.}}
 \label{tab:jhk}
 \end{table*}

Next, we corrected the mean magnitudes for extinction. As stated
in section~\ref{section:data}, we derived two extinction values 
from two different reddening maps. We obtained E(\jmk)$_{G12}$ and $A_{Ks(G12)}$ for each target 
from the online tool BEAM\footnote{\url{mill.astro.puc.cl/BEAM/calculator.php}}. 
However, estimating E(\jmk)$_{S14}$ is not straightforward
because the distance of the target needs to be known in order to obtain E(\jmk)$_{S14}$.
Distance and reddening in S14 are degenerate. 
Therefore, we adopted the following method. First, we located the 
four pixels of the map that are closest to the position of our target.
We weighted the E(\jmk)$_{S14}$ of each pixel by their inverse angular 
distance of the pixel center from our target. We repeated this for each of the 21 bins and obtained 
21 possible values for E(\jmk)$_{S14}$ for each target. Using an iterative method
\citep{bhardwaj17c}, we simultaneously found the most plausible 
values of E(\jmk)$_{S14}$ and $A_{Ks(S14)}$ and an estimate of the distance.

We adopted a new reddening law with both reddening maps to derive the extinction 
$A_{Ks}$ from E(\jmk). This law was derived 
by \citet{alonsogarcia2017} using VVV data for the innermost regions
of the bulge between $|l|$<2\Deg7 and $|b|$<1\Deg55.
Specifically, we adopted as the ratio of total-to-selective
extinction $R_{JK}=\dfrac{A_{Ks}}{E(\jmk)}$ the values 
shown in their Table 2, according to the quadrant, 
within their surveyed sky area. For targets outside the quoted area, we adopted $R_{JK}$=0.428,
which is the average suggested by \citet{alonsogarcia2017}.

We checked the differences $\Delta A_{Ks}$ = $A_{Ks(G12)} - A_{Ks(S14)}$
for all our targets, and found that $\Delta A_{Ks}$ values follow an almost
Gaussian distribution with a mean of --0.007 mag and $\sigma$=0.032 mag.
A tail of targets with 0.1<$\Delta A_{Ks}$<0.25 mag and one target with $\Delta A_{Ks}$
as high as 0.63 mag were found. Almost all of these targets are located at 
distance moduli (DM) smaller than 14.5 mag ($\sim$8 kpc). This means that they are closer than the Galactic
center. On the other hand, targets with $\Delta A_{Ks}$<--0.1 mag are 
mostly located at DMs larger than 14.5 mag. This is  
expected when comparing a reddening map fixed at $\sim$8 kpc (G12)
with one that takes distance into account (S14). This also suggests 
that we can adopt the S14 map to derive our final results.

However, since reddening and distance were 
derived simultaneously, $A_{Ks(S14)}$ may depend on the calibration 
selected for the PL relation. We checked that for different
calibrations (see Section~\ref{section:dist}), the differences between 
$A_{Ks(S14)}$ from different calibrations are within 0.05 mag and 
are smaller than 0.01 mag for $\sim$80\% of the targets.





\section{PL relations and distances}\label{section:pls}

\subsection{Empirical PL relation}\label{section:pl_empirical}

\begin{figure*}[!htbp]
\centering
\includegraphics[width=11cm]{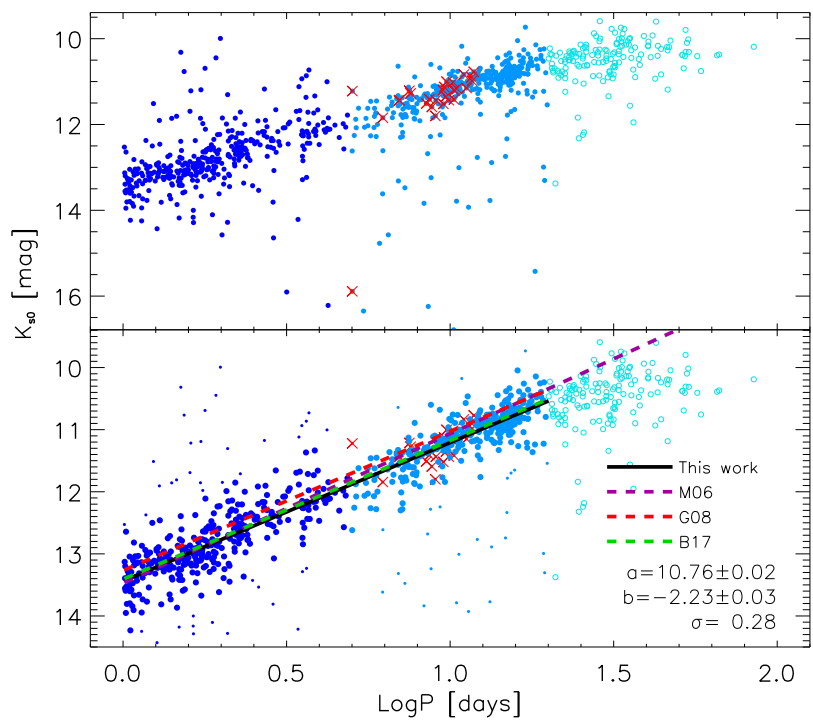}
\vspace{-0.2cm}\caption{Top: Bulge T2Cs in the period-luminosity plane.
Dark blue, light blue, and cyan circles mark BLHs, WVs, and RVTs, respectively.
Red crosses mark pWVs. Bottom: Close-up of the empirical PL$K_{s0}$
relation obtained by fitting BLHs and WVs in our sample (black solid line). 
Larger symbols display BLHs and WVs that were kept after an iterative 3$\sigma$ 
clipping procedure. The purple, red, and green 
dashed lines display the literature PL$K_{s0}$ shifted by 14.6 mag ($\sim$8.3 kpc) of
\citet{matsunaga06}, \citet{groenewegen08}, and \citet{bhardwaj17b}, respectively.}
\label{fig:pl}
\end{figure*}

Figure~\ref{fig:pl} shows the targets in the $\log{P}$-$K_{s0}$ plane.
The BLHs and WVs were dereddened adopting the S14 map, but $A_{Ks(S14)}$
values and distances were estimated simultaneously. Since we did not estimate the 
distances of pWVs and RVTs, as explained below, these targets 
were dereddened with $A_{Ks(G12)}$, which is independent of distance.

In stellar system like the bulge, where stars are not all
at the same distance (as in the case of
globular clusters, external galaxies, etc.), extinction and distance
are degenerate and a simple empirical PL does not provide
precise insight into the structure, especially if the reddening is not constant.
Moreover, there is a debate on whether the old population in the bulge
is indeed spheroidal \citep{dekany2013,kunder2016} or 
if it is ellipsoidal and tilted, similarly to the Galactic bar \citep{pietrukowicz2015}.
Therefore, as done before by \citet{groenewegen08}, we fit the PL 
relation by adding the dependence on the Galactic longitude ($l$) 
and latitude ($b$). We selected all the BLHs and WVs from our sample, and 
after an iterative rejection of outliers at more than 3$\sigma$, we found 
$K_{s0} = (10.66\pm0.02) - (2.21\pm0.03)\cdot(\log{P}-1.2) - (0.020\pm0.003)\cdot l + (0.050\pm0.008)\cdot |b|$
mag, with a standard deviation of 0.07 mag. The positive coefficient in $|b|$ 
means that fainter stars are located at higher distances from the Galactic plane,
where the reddening is lower. This is an evidence that the T2C sample is 
biased by reddening. On the other hand, the non-zero dependence of $K_{s0}$ 
on $l$ indicates that the T2C ellipsoid is tilted.
If we ignore the $l$ and $|b|$ terms, the simple PL relation is 
$K_{s0} = (10.76\pm0.02) - (2.23\pm0.03)\cdot(\log{P}-1.2)$ mag, and the
standard deviation increases to 0.28 mag. 
Figure~\ref{fig:pl} shows our newly derived empirical, 
coordinate-independent PL as a black solid 
line. We set the 
zero of the independent variable at $\log{P}$=1.2 
to facilitate comparison with \citet{groenewegen08}.
The coefficients of both the coordinate-dependent 
and the simple PL relations agree remarkably well with 
those obtained with an identical approach by
\citet{groenewegen08}: all of them agree within 
1$\sigma$. However, while the error on the coefficient
of the $l$ coordinate in \citet{groenewegen08} was larger than the value
itself (--0.028$\pm$0.031 mag/\deg, thus including the zero value 
within 1$\sigma$), our coefficient is more precise (--0.019$\pm$0.003 mag/\deg) because the set of variables is much larger, and it clearly 
rules out a PL relation that is independent of $l$.
We also compared our empirical
PL with that of \citet[][ $K_{s0} = (10.749\pm0.0056) - 
(2.189\pm0.032)\cdot(\log{P}-1.2)$ mag]{bhardwaj17c}. We adjusted their
zero-point since they had adopted $\log{P}$=1.0 as the zero of their
independent variable. The slope and zero-point both agree with
ours within 1$\sigma$.

The bottom panel of Figure~\ref{fig:pl} also shows PLs from the literature,
Galactic globular clusters (\citealt{matsunaga06}), bulge (\citealt{groenewegen08}), 
Large Magellanic Cloud (LMC; \citealt{bhardwaj17b}) 
as purple, red, and green dashed lines, respectively.
The PL relations of \citet{matsunaga06} and \citet{bhardwaj17b} were
placed at a DM of 14.6 mag ($\sim$8.3 kpc), which is the best recommended
value of $R_0$ in the literature in general \citep{blandhawthorn2016,degrijs2016}
and a very common value obtained from methods based on the 
PL of RRLs and T2Cs \citep{dekany2013,pietrukowicz2015,bhardwaj17c,majaess2018}.
We note that the slope of our relation is the same within
the uncertainties as the slope found by 
\citet[][b$_{B17}$=--2.21$\pm$0.02 mag]{bhardwaj17b}.

Finally, we note that RVTs are systematically fainter
than predicted by the PLs, which would be $K_{s0}$<10 mag. 
This is an effect of saturation, which in the
VVV is severe at magnitudes brighter than $K_s\sim$10.0 mag
\citep{minniti2010,mauro2013}. This is also supported by
the fact that as we discuss in Section~\ref{section:dist}, RVTs 
are expected to be either brighter than predicted by the PL 
\citep{matsunaga2009a,matsunaga11a,ripepi2015} or to 
follow the PL \citep{matsunaga06,bhardwaj17b}, but not to be fainter.
We also checked for possible selection biases that 
would favor the selection of the closest RVTs, but found none. This 
further supports the saturation scenario for these variables.

\subsection{Individual distances}\label{section:dist}

Before deriving individual distances, we discuss a few key points below.

{\it 1) Metallicity dependence.} There is a general consensus about 
the independence, or a very mild dependence, less
than 0.1 mag/dex \citep{dicriscienzo07}, of the PL relation of T2Cs on 
metal abundance, based on both empirical 
\citep{matsunaga06,matsunaga11a} and theoretical 
\citep{bono97e,dicriscienzo07} arguments.
The advantage of adopting a metal-independent PL relation to calibrate
distances is straightforward because the metallicity of our targets is
unknown. Since T2Cs belong to the same stellar population
as RRLs, their metallicity dispersion should be similar, and it 
should be fairly small \citep[0.25 dex,][]{pietrukowicz2012}.

{\it 2) RV Tauri and peculiar W Virginis.}  
T2Cs are separated into BLHs (1 day<P<5 days), 
WVs (5 days<P<20 days), and RVTs (P>20 days). The thresholds are those
by \citet{soszynski2011}. They are based on the period distribution of 
OGLE III T2Cs, and the authors kept them unchanged in the OGLE 
IV catalog \citet{soszynski2017}.
However, there is a debate as to whether 
RVTs obey the same PL as shorter-period T2Cs: \citet{matsunaga2009a} using data  from the InfraRed Survey Facility (IRSF) and \citet{ripepi2015} using VMC data, showed that 
in the LMC, RVTs are overluminous with respect to the 
extrapolation at long periods of the PL of BLHs and 
WVs. In contrast, \citet{bhardwaj17b} used NIR 
data in the central bar of the LMC \citep{macri2015} and found that RVTs also fall on the linear
PL fit to short-period T2Cs in the LMC. However, their photometry for short-period
variables (that is, at fainter magnitudes) was more prone to be affected
by crowding in the central regions than longer-period variables. 
On the other hand, RVTs in Globular clusters also follow the same PL relation
as shorter period T2Cs \citep{matsunaga06}. 
We do not discuss this in detail since this is not the aim of this 
work, but it is crucial to remember that in addition to these empirical
findings, there are two different
evolutionary channels from which RVTs are generated: either from 
low-mass ($\sim$0.50 M$_{\odot}$), very old (>10 Gyr) AGB stars 
\citep{wallerstein2002}, or from more massive ($\gtsim$1 M$_{\odot}$) 
and younger objects \citep{willsontempleton2009}.
Recent findings about RVTs also involve binarity, to
distinguish their evolutionary channel \citep{manick2018}.
This means that RVTs are
not reliable as distance indicators, and we did not 
take them into account for our PL relations. Finally, following the 
classification in \citet{soszynski2017}, we also discarded pWVs, 
which are a subclass of WVs that likely belong to 
binary systems \citep{soszynski08c}. They are overluminous 
\citep[0.3-0.5 mag in the $K_s$ band][]{ripepi2015} when compared to WVs with similar
periods. This means that they do not follow the PL relation of T2Cs and cannot
be used as distance indicators. We note that pWVs in Figure~\ref{fig:pl}
are not more luminous than WVs at the same period. Nonetheless, we
did not take them into account for the quoted reasons.

{\it 3) No semi-empirical calibration.}  In principle, it is not possible, 
in our case, to adopt a semi-empirical calibration (empirical slope from
our own sample and zero-point from literature) for the PLs because our targets are not 
all located at the same distance. This is true
in principle, even though our empirical slope is identical within 0.01 mag to that by
\citet{bhardwaj17b}, as stated in section~\ref{section:pl_empirical}.


{\it 4) Selection of the calibration.} Keeping these points in mind, we have searched the literature, 
where several calibrations of the PL
relation of T2Cs in the NIR are available 
\citep[][henceforth PL$_{M06}$,PL$_{D07}$,PL$_{M09}$,PL$_{R15}$,PL$_{B17}$ and PL$_{C17}$]{matsunaga06,dicriscienzo07,matsunaga2009a,ripepi2015,bhardwaj17b,clementini2017}.
We rule out the preliminary calibration PL$_{C17}$, based on {\it Gaia} DR1 
parallaxes, because it yields DM$_{LMC}$>18.9 mag, which does not 
agree with literature values nor with the {\it Gaia} DR1 calibration of RRLs and 
Classical Cepheids derived in the same work.
We note that PL$_{M09}$, PL$_{R15}$ and PL$_{B17}$, which 
are all based on T2Cs in the LMC, do not provide absolute calibrations of their PLs. 
Therefore, we adopted the distance of the LMC 
based on eclipsing binaries \citep[][DM$_{LMC}$=18.493$\pm$0.008$\pm$0.047 mag]{pietrzynski2013}
to set the zero-point. In principle, we could calibrate the zero-point using 
Baade-Wesselink parallaxes for field T2Cs obtained by \citet{feast08}, but 
this would be based only on two objects, with a strong 
difference among the different values of the parallaxes in the literature.
After checking each of these calibrating PLs, we decided to adopt PL$_{B17}$.
The choice was guided by the following reasons. First, the calibrating 
PL must be based on both BLHs and WVs. This rules out PL$_{D07}$
since they only used BLHs pulsation models. Second, of the three PLs from the 
LMC (PL$_{M09}$, PL$_{R15}$ , and PL$_{B17}$), the latter is based on
light curves with an average of 50 epochs, which is more than twice 
that of the other two together. Moreover, for targets that are outside their 
surveyed sky area, they include data from the previous works. 
Third, we excluded the M06 calibration because it is based on 
all the three subclasses of T2Cs, including RVTs, which we did not
include in our distance analysis. In the end, we adopted PL$_{B17}$, 
which is based only on BLHs and WVs, which are the subclasses of
variables for which we did estimate the distance.


As stated in section~\ref{section:data}, we simultaneously
estimated distances and the 3D extinction E(\jmk)$_{S14}$. This was made iteratively, 
following the method of \citet{bhardwaj17c}. Normally, after the 
second or third iteration, the values of distance and reddening converge.
We derived individual distances for 710 variables, which are
listed in Table~\ref{tab:dist}. Taking all sources of 
uncertainty into account (uncertainty on the mean magnitude, on the extinction, 
on the coefficients of the calibrating PL, and the intrinsic width of the PL), 
the relative uncertainties are in the range 8\%-9\% for 676 objects, 9\%-20\% for 38, and only one, at 
a distance of 2.87 kpc, has a relative uncertainty of $\sim$27\%.

\subsection{Spatial distribution of Type II Cepheids}\label{section:distribution}

Knowing the coordinates ($l$,$b$) and the distances $d$, 
we derived the coordinates $x_{GAL}$, $y_{GAL}$
and $z_{GAL}$. We adopted a reference frame with the Sun at the origin 
($x_{GAL}$=$y_{GAL}$=$z_{GAL}$=0), $x_{GAL}$ which increases toward
the Galactic center, $y_{GAL}$ which is on the Galactic plane and positive for $l$>0, and 
$z_{GAL}$ which is perpendicular to the Galactic plane and positive for $b$>0.

\begin{table*}[!htbp]
 \footnotesize
 \caption{Extinction, distances, and Cartesian coordinates of target T2Cs.}
  \centering
 \begin{tabular}{l c c c c c c c c c}
 \hline\hline  
ID & $A_{Ks(G12)}$\tablefootmark{a} & $A_{Ks(S14)}$ & $x_{GAL}$ & $y_{GAL}$ & $z_{GAL}$ & $d$ \\
  & mag & mag & kpc & kpc & kpc & kpc \\
 \hline 
 001 &     \ldots      &     \ldots    &   \ldots       &    \ldots        &    \ldots       &     \ldots      \\
 002 &   0.07$\pm$0.04 & 0.11$\pm$0.01 &  7.44$\pm$0.62 & --1.06$\pm$0.09  &  0.62$\pm$0.05  &  7.54$\pm$0.63  \\
 003 &   0.12$\pm$0.04 & 0.16$\pm$0.01 &  9.09$\pm$0.77 & --1.23$\pm$0.10  &  0.70$\pm$0.06  &  9.20$\pm$0.78  \\
 004 &     \ldots      &     \ldots    &   \ldots       &    \ldots        &    \ldots       &    \ldots       \\
 005 &   0.17$\pm$0.04 & 0.21$\pm$0.01 & 19.25$\pm$1.65 & --2.20$\pm$0.19  &  1.30$\pm$0.11  & 19.42$\pm$1.66  \\
 006 &     \ldots      &     \ldots    &   \ldots       &    \ldots        &    \ldots       &    \ldots       \\
 007 &   0.21$\pm$0.05 & 0.25$\pm$0.03 &  9.15$\pm$0.78 & --0.43$\pm$0.04  &  0.63$\pm$0.05  &  9.18$\pm$0.79  \\
 008 &   0.17$\pm$0.04 & 0.20$\pm$0.03 & 10.55$\pm$0.91 & --0.31$\pm$0.03  &  0.83$\pm$0.07  & 10.58$\pm$0.91  \\
 009 &   0.20$\pm$0.05 & 0.21$\pm$0.03 &  9.36$\pm$0.80 & --0.48$\pm$0.04  &  0.59$\pm$0.05  &  9.39$\pm$0.80  \\
 010 &     \ldots      &     \ldots    &   \ldots       &    \ldots        &    \ldots       &     \ldots      \\ 
 \hline
 \end{tabular}
 \tablefoot{Only the first 10 of the 924 lines of the table are shown. The full table is
 shown in the machine-readable version of the paper.\\
  \tablefoottext{a}{For some coordinates, the G12 map does not provide
an  error on E(\jmk), therefore there is no error on $A_{Ks(G12)}$.}}
\label{tab:dist}
 \end{table*}
 
\begin{figure*}[!htbp]
\centering
\includegraphics[width=10cm]{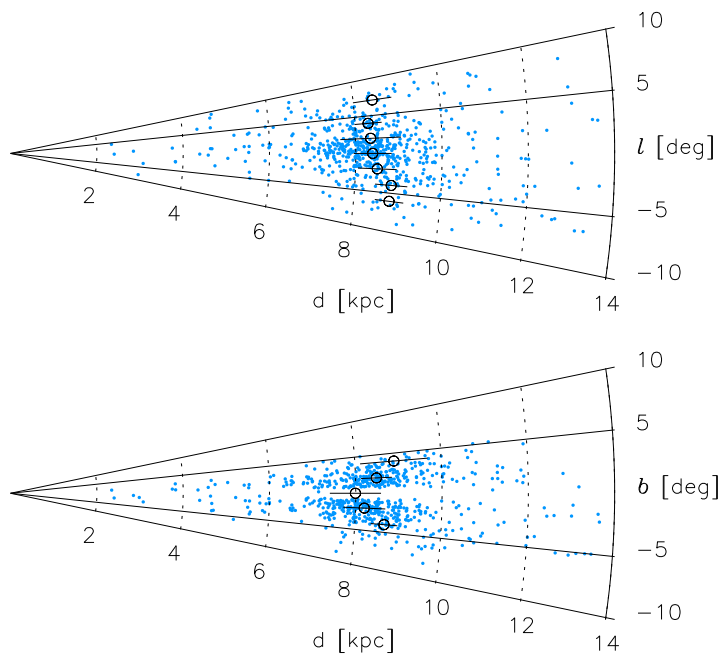}
\vspace{-0.5cm}\caption{Top: Projection, on the 
Galactic plane, of T2Cs within 14 kpc. The Sun is at the
vertex of the plot. Black circles with error bars represent the 
means and standard deviations of the distances projected onto the
Galactic plane, in 2{\deg} wide angular areas (4{\deg} wide
for the most peripheral one at positive $l$). Thirty-four T2Cs were not plotted because
they are more distant than 14 kpc. Bottom: Same as the top panel, 
but projected in latitude. The angular areas are 2{\deg} wide, except 
the central one, which is 3{\deg} wide.}
\label{fig:map}
\end{figure*}

Figure~\ref{fig:map} clearly shows that several objects are located either 
in front of or beyond the bulge. The 
individual distances range from 2.0 to 111.7 kpc, 
with 11 objects more distant than 30 kpc, most likely belonging
to the outer halo, which dominates at Galactocentric distances
greater than $\sim$20 kpc \citep{carollo07,carollo2018}.

The two panels of Figure~\ref{fig:map} also show as black circles
with error bars the means and standard deviations of the distances 
projected onto the Galactic plane (top panel) or onto the longitude ($l$=0) plane
(bottom panel), in angular areas of 2{\deg}.
We point out that in the top panel, the most peripheral area at 
positive $l$ is 4{\deg} wide, 
and in the bottom panel, the most central area is 3{\deg} wide, 
to take into account the lower density of objects.
These mean and standard deviations were derived by fitting 
a rescaled histogram with a Gaussian. A detailed explanation
is provided below. 

{\it Galactic ($b$=0) plane).}  The central angular areas
share basically identical averages. However, the most peripheral
areas (5\Deg0<|$l$|<9\Deg0, that is, between 0.7 and 1.25 kpc from the Galactic
center) show a slight deviation similar to that of the barred distribution
of Red Clump stars \citep{wegggerarhd2013}: closer at positive $l$ and 
more distant at negative $l$. This agrees with our finding that the PL relation
does depend on the $l$ coordinate, and gives further evidence of the 
ellipsoidal symmetry of the old stellar component of the bulge, 
as has been outlined by \citet{pietrukowicz2015}, who found an ellipsoidal
structure for bulge RRLs, with an inclination of the major axis with 
respect to the Sun of 20\deg$\pm$3\deg, similar to the orientation 
of the Galactic bar \citep[$\sim$30\deg][]{gonzalez2011,wegggerarhd2013}.

{\it Longitude ($l$=0) plane).}  A distinct yet fictitious trend 
of the average distance with $b$ is 
displayed in the bottom panel of Figure~\ref{fig:map}. 
The variables that are closer to the Galactic plane appear to have
smaller distances. This is due to a selection effect 
caused by extinction in the original OGLE catalog, 
which more easily detects stars in the closest part  
of the bulge than stars in the farthest part, which
are more heavily reddened. The extinction ratio between the $I$ band of OGLE and the $K_s$ band
of VVV ranges within a factor of four to ten, and 
the lower limit of $A_{Ks}(S14)$ for our targets is $\sim$0.3 mag. 
We have considered two other possible explanations for this 
trend: either an overestimate of reddening at low $b$ or 
the effect of crowding on aperture photometry, but none 
can explain the quoted behavior.
We discard the possibility that reddening is 
overestimated because other experiments that 
adopted either the G12 or the S14 map, which
are consistent between themselves, provided 
estimates of $R_0$, all at about 8.3 kpc 
\citep{gonzalez2012,bhardwaj17c,majaess2018}. We also rule out 
the possibility that crowding affects the magnitudes from
aperture photometry, making the targets brighter. A direct 
comparison of PSF versus aperture photometry does not 
reveal any clear trend with $b$, and the average difference
of mean magnitudes is $\Delta K_{s(Aperture-PSF)}$=0.03$\pm$0.013 mag.

With individual distances and coordinates for our targets, 
we can estimate the distance of the 
Galactic center $R_0$. However, the calculation is not
straightforward, and cuts and resampling are needed to 
take the biases into account. 

First, we selected only stars
at $R_{G}=\sqrt{x_{GAL}^2+y_{GAL}^2}$, which is the 
distance of the star, projected onto the Galactic plane, between
6 and 11 kpc, to avoid non-bulge stars within the
sample. The choice is justified by the results 
of \citet[][see their Fig. 5]{pietrukowicz2015}, who showed that the density of bulge stars is very low 
(lower than $\sim$10\% of the peak) outside this distance range.
Second, we only selected stars at $b$>3\Deg0 and $b$<--3\Deg0 to 
avoid the OGLE selection bias. A similar
cut was applied by \citet{pietrukowicz2015}, who only used RRLs at 
$b$<--2\Deg7 to estimate $R_0$. 

These selection criteria left us with 172 stars. 
However, their average longitude ($\langle l \rangle$) is
0\Deg275, which means that the sample is biased toward shorter 
distances. To overcome this bias, we performed a resampling of
the data by randomly selecting 75 stars at negative $l$
and 75 at positive $l$. If $\langle l \rangle$ of the 150
random targets is lower in absolute values than 0\Deg1, we 
kept the sample, otherwise, we repeated the random
target selection. We point out that this
process is not a proper bootstrap method because 
we did not allow sampling the same element more than once.

\begin{figure*}[!htbp]
\centering
\includegraphics[width=10cm]{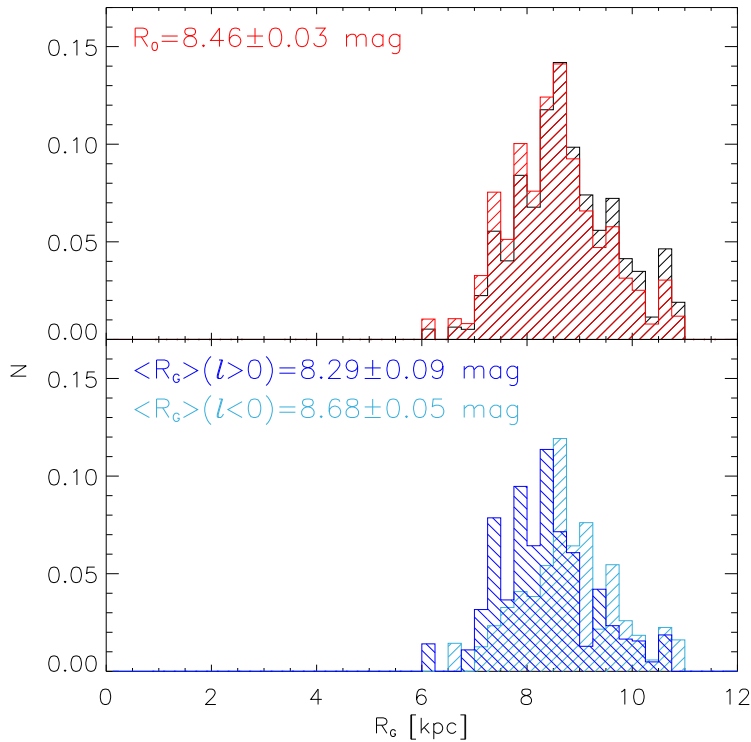}
\caption{Top: Overall histogram of the distances
of 150 targets resampled 5,000 times (750,000 in total, shown in black).
 $d^{-2}$ scaled histogram (red). A Gaussian centered
at $R_0$ is displayed. The estimate of $R_0$ is labeled 
with its uncertainty as derived by using 
percentiles, as described in the text.
Bottom: $d^{-2}$ scaled histogram of 
75 targets at $l$<0\deg resampled 5,000 times (375,000 in total, shown in blue)
Light blue: Same as the  blue, but for targets at $l$>0\deg.}
\label{fig:distance2}
\end{figure*}

With this set of 150 targets, we plot the distribution of $R_{G}$ 
in bins of 0.25 kpc as shown in Figure~\ref{fig:distance2} (black histogram).
However, this distribution is biased and shifted to greater distances. 
At fixed coordinates ($l$,$b$), the  
volume within a given sky area ($\Delta l$,$\Delta b$) 
and a given depth range ($\Delta d$) increases with 
distance. This means that the number of stars 
within the volume (and therefore the probability 
of detecting a target in the volume) increases quadratically with distance. 
This causes a bias that shifts the distribution toward the 
more probable larger distances.
To take this geometric effect into account, we scaled the 
distribution by $d^{-2}$. We fit the scaled distribution (red histogram in 
Figure~\ref{fig:distance2}) with a Gaussian. 
We estimated the abscissa of the peak ($x_0$) and adopted it as our estimate of $R_{0(i)}$ on the $i$-th resampled set. 
Starting from the random extraction
of 150 targets, this process was repeated 5,000 times to avoid any selection bias.

\begin{figure*}[!htbp]
\centering
\includegraphics[width=10cm]{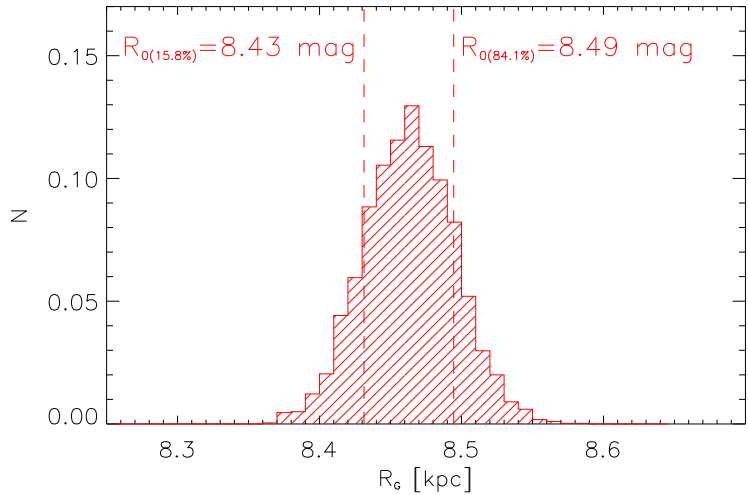}
\caption{Distribution of the 5,000 estimates of $R_0$ from
the resampled sets of targets. The dashed lines display the percentiles
at 15.8\% and at 84.1\% that we used to derive the uncertainty
on $R_0$.}
\label{fig:distance3}
\end{figure*}

We adopted an overall average of the 5,000 estimates of 
$R_{0(i)}$ as our final estimate of $R_0$. Based on a sample with 
$\langle l \rangle$=0\Deg016$\pm$0\Deg050,
we obtain a final $R_0$ estimate of 8.46 kpc. The statistical
uncertainty of both $R_0$ was 
derived as half of the range between the 
15.8\% and 84.1\% percentiles of the  
distribution of $R_{0(i)}$ (see Fig.~\ref{fig:distance3}). These
thresholds were chosen to enclose 68.3\% of the estimates 
provided by the simulations, like a $\pm$1$\sigma$ 
range in a Gaussian distribution. We derived a range of 8.43-8.49 kpc for 
$R_0$, which means a statistical uncertainty of 0.03 kpc.

We calculated the systematic
uncertainty as the squared sum of the average uncertainty on the 
mean magnitude (0.020 mag), the average uncertainty on the extinction (0.021 mag), and 
the average propagation of the uncertainties of the calibrating
PL coefficients (almost vanishing, 0.001 mag). These together are
0.028 mag, which is 0.11 kpc at 8.30 kpc. Our estimate of $R_0$ does not
agree very well with estimates from similar works, either
using T2Cs \citep[8.34$\pm$0.03{[}stat.{]}$\pm$0.41{[}syst.{]} kpc,][]{bhardwaj17c} or RRLs
(8.33$\pm$0.05{[}stat.{]}$\pm$0.14{[}syst.{]} kpc, \citealt{dekany2013}; 
8.27$\pm$0.01{[}stat.{]}$\pm$0.40{[}syst.{]} kpc, \citealt{pietrukowicz2015}). 
However, all the quoted papers adopted the reddening law by \citet{nishiyama2009}, 
which provides a higher $R_{JK}=0.528$ and, in turn, smaller distances. 
Had we adopted the \citet{nishiyama2009} reddening law with our data, it
would have provided $R_0$=8.30$\pm$0.03(stat.)$\pm$0.11(syst.), which would agree 
perfectly well with the quoted papers. This is evidence of 
how crucial a correct understanding of the reddening law is. Our estimate 
of $R_0$ agrees within 1$\sigma$ with the best overall recommended value from 
a recent review \citep[$\sim$8.3$\pm$0.2{[}stat.{]}$\pm$0.4{[}syst.{]} kpc,][]{degrijs2016}.

Finally, a more detailed analysis of the data allows us to 
show further evidence of the asymmetrical distribution of 
T2Cs around the Galactic center. By resampling 5,000 times
only T2Cs at positive $l$, we obtain an average peak of the distribution
of 8.29$\pm$0.09 kpc. The same process on T2Cs at negative $l$ provides a 
value of 8.68$\pm$0.05 kpc. Together with the distribution of average 
distances in the Galactic plane (top panel of Figure~\ref{fig:map}) 
and the dependence of the PL relation on $l$, this is strong evidence
that T2Cs trace an old, ellipsoidal stellar population.

\section{Kinematics}\label{section:kinematics}

\subsection{Proper motion of the center of mass}

The proper motion of Sgr A*, the supermassive black hole at the center of
the Milky Way, based on VLBA measures, is $\mu_{l*(Sgr A*)}$=--6.379$\pm$0.026 mas/yr; 
$\mu_{b(Sgr A*)}$=--0.202$\pm$0.019 mas/yr \citep{reid2004}. Assuming that the 
center of mass of the old population traced by T2Cs overlaps with Sgr A* and
has the same proper motion, we adopt the proper motions of T2Cs to 
obtain an indirect estimate of the proper motion of the center of mass.

As discussed in Section~\ref{section:data}, we collected 
proper motions from three different catalogs: 
VIRAC, PSF, and {\it Gaia}. For the 
analysis in this section, we rejected, from all three catalogs proper motions 
with a combined statistical error $CSE=\sqrt{err\mu_{\alpha *}^2+err\mu_{\delta}^2}$ or
$CSE=\sqrt{err\mu_{l*}^2+err\mu_{b}^2}$ larger than 2 mas/yr, leaving 553, 343, and 837
targets from VIRAC, PSF, and {\it Gaia}, respectively.
We point out that the error propagation for the {\it Gaia} proper motions, when converting
from ($\mu_{\alpha *}$,$\mu_{\delta}$) into ($\mu_{l*}$,$\mu_{b}$), 
was performed taking into account the covariance terms as suggested by \citet{gaia_dr2_parallax}.
 {\it Gaia} coordinates were precessed from their native J2015.5 epoch
to J2000, the same reference epoch as for VIRAC and PSF, to perform the quoted conversion.

\begin{figure*}[!htbp]
\centering
\includegraphics[width=12cm]{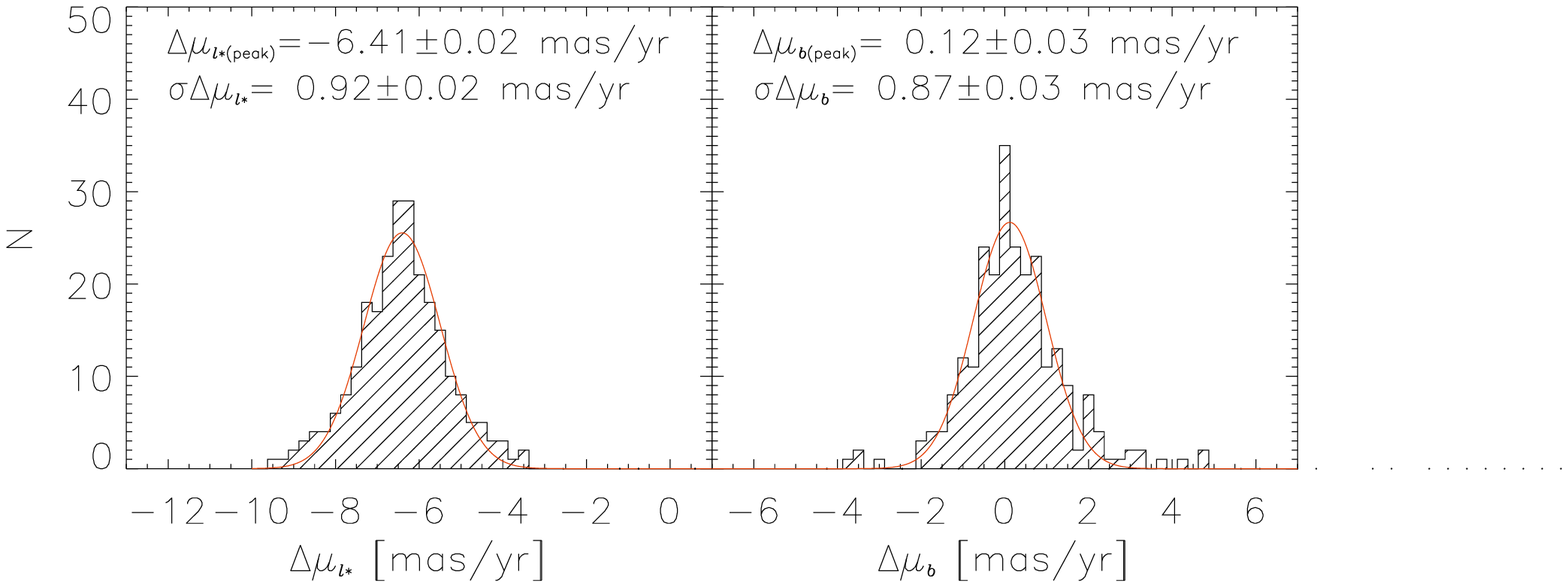}
\caption{Left: Distribution of $\Delta\mu_{l*}$ 
for 251 T2Cs located less than 2 kpc away from the center of the 
Galaxy. A Gaussian fit to the distribution is shown in red. The abscissa of
the peak and the $\sigma$ of the fit are labeled. Right: Same as left panel,
but for $\Delta\mu_{b}$.}
\label{fig:deltamu}
\end{figure*}

It is crucial to remember that while
VIRAC and PSF provide relative proper motions in the frame of reference
of the Galaxy, {\it Gaia} provides absolute proper motions in the practically inertial
frame of reference defined by quasars. This allows an interesting comparison
among the catalogs. As a first step, 
we left out VIRAC proper motions. As displayed in Figure~\ref{fig:deltamu}, the 
distribution of $\Delta\mu_{l*}=\mu_{l*(Gaia)}-\mu_{l*(PSF)}$
is centered at $\Delta\mu_{l*(peak)}$=--6.41$\pm$0.02 mas/yr. 
For the $b$ component, we find $\Delta\mu_{b(peak)}$=0.12$\pm$0.03 mas/yr. 
These numbers were derived using 251 T2Cs for which we have both
{\it Gaia} and PSF proper motions and that are located within 2 kpc from the
center of the Galaxy, as derived in Section~\ref{section:distribution}. The latter 
criterion was adopted as a compromise to leave out possible thick-disk stars and to
retain a large sample of targets.

When we assume that the velocities of our targets are randomly 
distributed around the center of mass, which is reasonable because this is the
behavior of old, low-mass bulge stars 
\citep{spaenhauer1992,minniti1996,babusiaux2010,kunder2016}, then 
($\Delta\mu_{l*(peak)}$,$\Delta\mu_{b(peak)}$)
is an indirect estimate of the proper motion of the center of mass of the Galaxy.
This is supported by the fact that $\Delta\mu_{l*(peak)}$ is identical well within 1$\sigma$ to
the longitudinal component of the proper motion of Sgr A*.
The latitudinal component $\Delta\mu_{b(peak)}$ does not agree with that of Sgr A*,
but the mean uncertainty on the proper motions (0.82 mas/yr for PSF and 0.47 for 
{\it Gaia}) is larger than the offset. We also checked whether there is a trend of $\mu_{l*}$ and
$\mu_{b}$ with distance, and found nothing significant in either 
the {\it Gaia} or the PSF sample. This implies that there is no evidence of net rotation.

We have performed the same analysis using 
proper motions from VIRAC and {\it Gaia} and 
found that in this case, $\Delta\mu_{l*(peak)}=-5.43\pm0.03$ mas/yr and 
$\Delta\mu_{l*(peak)}=-0.04\pm0.02$ mas/yr. While the $b$ component
is similar to that of Sgr A*, the component along $l$ is different from
$\mu_{l*(Sgr A*)}$ by as much as $\sim$1 mas/yr. 
Moreover, we  found that for 265 targets in common between
the VIRAC and PSF datasets, the difference between the medians of 
$\mu_{l*}$ and $\mu_{b}$, are 
--0.97$\pm$2.62 and +0.12$\pm$2.62 mas/yr, respectively. 
No dependence of $\mu_{l*}$ on $l$ or on $b$ was found
in either of the two catalogs. We conclude that the $\mu_{PSF}$
are more reliable than $\mu_{VIR}$ not only a priori, as discussed in Section~\ref{section:data},
but also a posteriori. Because these are both relative measures
of the proper motion, the assumption of randomly distributed motions
around the Galactic center would imply a zero median $\mu_{l*}$.
The comparisons with {\it Gaia} and that with the PSF sources both indicate that VIRAC
proper motions are systematically shifted by $\sim$1 mas/yr to the east
on the Galactic plane.

\subsection{Velocity dispersion}\label{section:veldispersion}

Tangential velocities can be used to test the kinematic properties
of a stellar population. If bulge T2Cs trace a spheroidal population 
that is kinematically hot, then their distribution should be dominated by a velocity 
dispersion with negligible rotation such as the Bulge RRL population \citep[][and references therein; Contreras Ramos et al. 2018, submitted]{minniti1996,kunder2016,marconiminniti2018}.
In this case, the velocity ellipsoid should be fairly symmetric in 
the Galactic longitude ($v_{l*}$) and latitude ($v_b$) components. 
To test this hypothesis, we derived the tangential velocity
($v_t$) and its two components $v_l*$ and $v_b$, 
using the classical relation $v_t = 4.74\cdot d \cdot \mu$, 
where $d$ is in kpc and and $\mu$ in mas/yr. We used $d$
as derived in Section~\ref{section:dist} and $\mu$ from
{\it Gaia} because it is the most complete sample with the
smallest uncertainties. While pWVs and RVTs might have valid 
values of $\mu$, we cannot derive their $v_t$, since $d$ is
not available. Table~\ref{tab:pm} displays the proper motions and 
velocities.

\begin{table*}
 \footnotesize
 \caption{Proper motions and velocities of target T2Cs.}
 \centering
 \begin{tabular}{l c c c c c c c c c}
 \hline\hline  
& \multicolumn{2}{c}{VIRAC} & \multicolumn{2}{c}{PSF} & \multicolumn{2}{c}{{\it Gaia}} & & & \\
 ID & $\mu_{l*}$ & $\mu_b$ & $\mu_{l*}$ & $\mu_b$ & $\mu_{l*}$ & $\mu_b$ & $v_l$\tablefootmark{a} & $v_b$\tablefootmark{a} & $v_t$\tablefootmark{a} \\
  & mas/yr & mas/yr & mas/yr & mas/yr & mas/yr & mas/yr & km/s & km/s & km/s \\
\hline
001 &         \ldots    &         \ldots    &           \ldots  &           \ldots  &   --8.91$\pm$0.09 &    0.01$\pm$0.12 &           \ldots  &        \ldots   &       \ldots    \\ 
002 &   --3.13$\pm$0.81 &    1.75$\pm$0.04 &           \ldots  &           \ldots  &   --8.49$\pm$0.06 &    2.50$\pm$0.06 &   --303.4$\pm$25.4 &   89.3$\pm$7.8  &  316.3$\pm$24.5  \\
003 &   --3.46$\pm$0.81 &   --1.05$\pm$0.06 &           \ldots  &           \ldots  &   --8.88$\pm$0.12 &   --2.03$\pm$0.11 & --387.4$\pm$33.2 & --88.5$\pm$8.9  &  397.3$\pm$32.5   \\
004 &         \ldots    &         \ldots    &           \ldots  &           \ldots  &   --2.27$\pm$0.09 &    2.92$\pm$0.08 &           \ldots  &        \ldots   &       \ldots     \\
005 &    2.63$\pm$1.48 &    1.68$\pm$0.06 &           \ldots  &           \ldots  &   --5.12$\pm$0.29 &   --0.47$\pm$0.26 &   --471.4$\pm$48.4 & --42.9$\pm$24.1 &  473.3$\pm$48.3   \\
006 &         \ldots    &         \ldots    &           \ldots  &           \ldots  &   --6.66$\pm$0.06 &   --0.90$\pm$0.07 &          \ldots  &        \ldots   &       \ldots      \\
007 &         \ldots    &         \ldots    &           \ldots  &           \ldots  &   --4.53$\pm$0.22 &   --4.54$\pm$0.14 & --197.0$\pm$19.5 &--197.4$\pm$18.1 &  278.9$\pm$18.8   \\
008 &    0.71$\pm$0.92 &   --1.25$\pm$0.12 &           \ldots  &           \ldots  &   --4.79$\pm$0.21 &   --0.95$\pm$0.18 &  --240.2$\pm$23.2 & --47.9$\pm$ 9.7 &  245.0$\pm$22.8   \\
009 &         \ldots    &         \ldots    &           \ldots  &           \ldots  &   --5.98$\pm$0.19 &   --1.67$\pm$0.18 & --266.3$\pm$24.2 & --74.2$\pm$10.1 &  276.5$\pm$23.4   \\
010 &         \ldots    &         \ldots    &           \ldots  &           \ldots  &   --5.21$\pm$0.10 &   --5.46$\pm$0.10 &          \ldots  &        \ldots   &       \ldots      \\    
\hline
 \end{tabular}
 \tablefoot{Only the first 10 of the 924 lines of the table are shown. The full table is 
shown in the machine-readable version of the paper.}
\tablefoottext{a}{The velocities are based on the absolute proper
motions by {\it Gaia}}
\label{tab:pm}
\end{table*}
 
\begin{figure*}[!htbp]
\centering
\includegraphics[width=12cm]{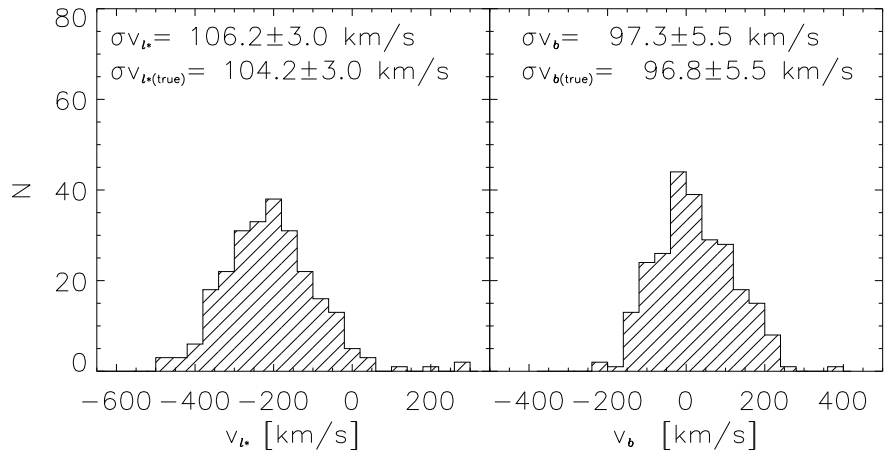}
\caption{Left: Distribution of the $l$ component of $v_t$ for 164 
T2Cs located at less than 2 kpc from the center of the 
Galaxy, as in Section~\ref{section:distribution}. The simple and true standard deviations
of the sample are labeled. Right: Same as left panel, but for the $b$ component
of $v_t$.}
\label{fig:vdispersion}
\end{figure*}

Finally, we derived the standard deviation of $v_{l*}$ and $v_b$ and obtained 
106.2$\pm$3.0 km/s and 97.3$\pm$5.5 km/s, respectively. These values were corrected by 
subtracting, in quadrature, the average uncertainties on $v_{l*}$ (20.5 km/s) and $v_b$ (9.9 km/s).
After this correction, we find $\sigma v_{l*}$=104.2$\pm$3.0 km/s,
and $\sigma v_b$=96.8$\pm$5.5 km/s. The agreement between the two is 
better than 1.2$\sigma$, thus providing further evidence that the T2Cs 
belong to a kinematically hot population. We cannot rule
out the possibility that despite our cuts, we still included some 
thick-disk objects. Stricter selections concerning the 
target distance from the center are hampered by the sample size. 



\section{Conclusions}\label{section:conclusions}

We have retrieved $K_s$-band light curves from VVV aperture photometry for 894 of 924 T2Cs in the 
OGLE IV catalog \citep{soszynski2017}. We calculated mean magnitudes
and amplitudes based on PLOESS fits \citep{persson2004,braga2018} to the light curves.
For BLHs and WVs, we  simultaneously estimated individual extinctions and distance moduli, based on 
a 3D reddening map \citep{schultheis2014} and on a PL relation. The calibration of 
the PL relation was based on the slope and zero-point of T2Cs in the LMC \citep{bhardwaj17b}, 
anchored with a late-type eclipsing binaries distance to the LMC \citep{pietrzynski2013}.
We found distances ranging from 2.0 to 111.7 kpc kpc, which means that our objects are located 
in the bulge, in the inner and outer halo, and possibly in the thick disk. The  mean individual 
relative uncertainty is 8.6\%, independent of distance and with a small standard 
deviation of 1.2\%.

The distribution of the individual distances, taking various geometric and selection biases into account, provides
an estimate of the distance of the 
Galactic center $R_0$ of 8.46$\pm$0.03(stat.)$\pm$0.11(syst.), which 
agrees with the recommended value of 8.3$\pm$0.2(stat.)$\pm$0.4(syst.) kpc 
\citep{degrijs2016}. Our estimate of $R_0$ does not agree with other 
estimates with similar methods 
\citep[$R_0 \approx 8.30$ kpc][]{dekany2013,pietrukowicz2015,bhardwaj17c},
but the difference is consistent with the different reddening law that was 
adopted (\citealt{alonsogarcia2017} instead of \citealt{nishiyama2009}).

We provided solid evidence that the old stellar population in the bulge is 
ellipsoidal. First, we found a non-negligible
dependence of the PL relation on the $l$ coordinate. This has been described before by \citet{groenewegen08}, but their limited sample hampered
the precision of the coefficient (--0.028$\pm$0.031 mag/\deg), while ours
is more precise (--0.019$\pm$0.003 mag/\deg). Second, we found that at 
$l\lesssim$--5\deg, the average distance is larger, while at $l\gtrsim$5\deg
the average distance of T2Cs is smaller, on a map projected onto the 
Galactic plane. Third, which is a similar but more quantitative approach as 
the second point, we found that the distribution 
of T2Cs at positive $l$ is centered at 8.29$\pm$0.09 kpc, while that 
of T2Cs at negative $l$ is centered at 8.68$\pm$0.05 kpc.

We also adopted proper motions from both {\it Gaia} and VVV itself to constrain 
the kinematic properties of T2Cs in the bulge. The analysis was restricted to only
the sources with a combined statistical error smaller than 2 mas/yr. The power of the synergy 
between {\it Gaia} and VVV astrometric data is clear when comparing the absolute proper
motions from {\it Gaia} with relative proper motions from the VVV. The mean difference 
(--6.41$\pm$0.02 and 0.12$\pm$0.03 mas/yr in the longitude and latitude 
direction, respectively) for T2Cs within 2 kpc from the Galactic center 
is similar within the uncertainties (0.82 mas/yr for PSF and 
0.47 mas/yr for {\it Gaia}) to the VLBA estimate of the relative proper 
motion of Sgr A* (--6.379$\pm$0.026 and --0.202$\pm$0.019 mas/yr). This is 
reasonable if we assume that the T2Cs of the bulge belong to the kinematically hot, 
old stellar population \citep{minniti1996,kunder2016}. Another piece of evidence supporting
this assumption is that the velocity dispersion in both the longitude 
and latitude directions agree within almost 1$\sigma$ 
($\sigma v_{l*}$=104.2$\pm$3.0 km/s, $\sigma v_b$=96.8$\pm$5.5 km/s.)
The difference may be due to contamination by thick-disk stars in the 
2 kpc sphere around the Galactic center.

It is important to note that while the distribution and kinematics of metal-rich 
populations in the bulge, tracing the X-shaped structure, have been studied widely, 
the distributions of more metal-poor populations based on different tracers remain to be 
investigated in detail. This work on T2Cs provides results that 
are consistent with RRLs. The spectroscopic follow-up of these objects 
in the near future will allow us to confirm the differences in their spatial distributions
and kinematics to those of metal-rich populations in the Galactic bulge.

\begin{acknowledgements}
We are grateful to the anonymous referee for the useful 
suggestions that helped us to improve our paper.
This work has made use of data from the European Space Agency (ESA) mission
{\it Gaia} (\url{https://www.cosmos.esa.int/gaia}), processed by the {\it Gaia}
Data Processing and Analysis Consortium (DPAC,
\url{https://www.cosmos.esa.int/web/gaia/dpac/consortium}). Funding for the DPAC
has been provided by national institutions, in particular the institutions
participating in the {\it Gaia} Multilateral Agreement.
\end{acknowledgements}

\bibliographystyle{aa}
\bibliography{../../../Latex/ms}

\begin{thebibliography}{63}
\expandafter\ifx\csname natexlab\endcsname\relax\def\natexlab#1{#1}\fi

\bibitem[{{Alonso-Garc{\'{\i}}a} {et~al.}(2017){Alonso-Garc{\'{\i}}a},
  {Minniti}, {Catelan}, {Contreras Ramos}, {Gonzalez}, {Hempel}, {Lucas},
  {Saito}, {Valenti}, \& {Zoccali}}]{alonsogarcia2017}
{Alonso-Garc{\'{\i}}a}, J., {Minniti}, D., {Catelan}, M., {et~al.} 2017, \apjl,
  849, L13

\bibitem[{{Baade}(1944)}]{baade44}
{Baade}, W. 1944, \apj, 100, 137

\bibitem[{{Babusiaux} {et~al.}(2010){Babusiaux}, {G{\'o}mez}, {Hill}, {Royer},
  {Zoccali}, {Arenou}, {Fux}, {Lecureur}, {Schultheis}, {Barbuy}, {Minniti}, \&
  {Ortolani}}]{babusiaux2010}
{Babusiaux}, C., {G{\'o}mez}, A., {Hill}, V., {et~al.} 2010, \aap, 519, A77

\bibitem[{{Bhardwaj} {et~al.}(2017{\natexlab{a}}){Bhardwaj}, {Macri},
  {Rejkuba}, {Kanbur}, {Ngeow}, \& {Singh}}]{bhardwaj17b}
{Bhardwaj}, A., {Macri}, L.~M., {Rejkuba}, M., {et~al.} 2017{\natexlab{a}},
  \aj, 153, 154

\bibitem[{{Bhardwaj} {et~al.}(2017{\natexlab{b}}){Bhardwaj}, {Rejkuba},
  {Minniti}, {Surot}, {Valenti}, {Zoccali}, {Gonzalez}, {Romaniello}, {Kanbur},
  \& {Singh}}]{bhardwaj17c}
{Bhardwaj}, A., {Rejkuba}, M., {Minniti}, D., {et~al.} 2017{\natexlab{b}},
  \aap, 605, A100

\bibitem[{{Bland-Hawthorn} \& {Gerhard}(2016)}]{blandhawthorn2016}
{Bland-Hawthorn}, J. \& {Gerhard}, O. 2016, \araa, 54, 529

\bibitem[{{Bono} {et~al.}(1997){Bono}, {Caputo}, \& {Santolamazza}}]{bono97e}
{Bono}, G., {Caputo}, F., \& {Santolamazza}, P. 1997, \aap, 317, 171

\bibitem[{{Braga} {et~al.}(2018){Braga}, {Stetson}, {Bono}, {Dall'Ora},
  {Ferraro}, {Fiorentino}, {Iannicola}, {Marconi}, {Marengo}, {Monson},
  {Neeley}, {Persson}, {Beaton}, {Buonanno}, {Calamida}, {Castellani}, {Di
  Carlo}, {Fabrizio}, {Freedman}, {Inno}, {Madore}, {Magurno}, {Marchetti},
  {Marinoni}, {Marrese}, {Matsunaga}, {Minniti}, {Monelli}, {Nonino},
  {Piersimoni}, {Pietrinferni}, {Prada-Moroni}, {Pulone}, {Stellingwerf},
  {Tognelli}, {Walker}, {Valenti}, \& {Zoccali}}]{braga2018}
{Braga}, V.~F., {Stetson}, P.~B., {Bono}, G., {et~al.} 2018, \aj, 155, 137

\bibitem[{{Carollo} {et~al.}(2007){Carollo}, {Beers}, {Lee}, {Chiba}, {Norris},
  {Wilhelm}, {Sivarani}, {Marsteller}, {Munn}, {Bailer-Jones}, {Fiorentin}, \&
  {York}}]{carollo07}
{Carollo}, D., {Beers}, T.~C., {Lee}, Y.~S., {et~al.} 2007, \nat, 450, 1020

\bibitem[{{Carollo} {et~al.}(2018){Carollo}, {Tissera}, {Beers}, {Gudin},
  {Gibson}, {Freeman}, \& {Monachesi}}]{carollo2018}
{Carollo}, D., {Tissera}, P.~B., {Beers}, T.~C., {et~al.} 2018, ArXiv e-prints,
  ApJL accepted [\eprint[arXiv]{1803.05154}]

\bibitem[{{Cole}(1998)}]{cole1998}
{Cole}, A.~A. 1998, \apjl, 500, L137

\bibitem[{{Contreras Ramos} {et~al.}(2017){Contreras Ramos}, {Zoccali},
  {Rojas}, {Rojas-Arriagada}, {G{\'a}rate}, {Huijse}, {Gran}, {Soto},
  {Valcarce}, {Est{\'e}vez}, \& {Minniti}}]{contreras2017}
{Contreras Ramos}, R., {Zoccali}, M., {Rojas}, F., {et~al.} 2017, \aap, 608,
  A140

\bibitem[{{de Grijs} \& {Bono}(2016)}]{degrijs2016}
{de Grijs}, R. \& {Bono}, G. 2016, \apjs, 227, 5

\bibitem[{{D{\'e}k{\'a}ny} {et~al.}(2013){D{\'e}k{\'a}ny}, {Minniti},
  {Catelan}, {Zoccali}, {Saito}, {Hempel}, \& {Gonzalez}}]{dekany2013}
{D{\'e}k{\'a}ny}, I., {Minniti}, D., {Catelan}, M., {et~al.} 2013, \apjl, 776,
  L19

\bibitem[{{Di Criscienzo} {et~al.}(2007){Di Criscienzo}, {Caputo}, {Marconi},
  \& {Cassisi}}]{dicriscienzo07}
{Di Criscienzo}, M., {Caputo}, F., {Marconi}, M., \& {Cassisi}, S. 2007, \aap,
  471, 893

\bibitem[{{Feast} {et~al.}(2008){Feast}, {Laney}, {Kinman}, {van Leeuwen}, \&
  {Whitelock}}]{feast08}
{Feast}, M.~W., {Laney}, C.~D., {Kinman}, T.~D., {van Leeuwen}, F., \&
  {Whitelock}, P.~A. 2008, \mnras, 386, 2115

\bibitem[{{Gaia Collaboration} {et~al.}(2018){Gaia Collaboration}, {Brown, A.
  G. A.}, {Vallenari, A.}, {Prusti, T.}, {de Bruijne, J. H. J.}, \& {et
  al.}}]{gaia_dr2}
{Gaia Collaboration}, {Brown, A. G. A.}, {Vallenari, A.}, {et~al.} 2018, A\&A

\bibitem[{{Gaia Collaboration} {et~al.}(2017){Gaia Collaboration},
  {Clementini}, {Eyer}, {Ripepi}, {Marconi}, {Muraveva}, {Garofalo}, {Sarro},
  {Palmer}, {Luri}, \& et~al.}]{clementini2017}
{Gaia Collaboration}, {Clementini}, G., {Eyer}, L., {et~al.} 2017, \aap, 605,
  A79

\bibitem[{{Gaia Collaboration} {et~al.}(2016){Gaia Collaboration}, {Prusti},
  {de Bruijne}, {Brown}, {Vallenari}, {Babusiaux}, {Bailer-Jones}, {Bastian},
  {Biermann}, {Evans}, \& et~al.}]{gaia_alldr}
{Gaia Collaboration}, {Prusti}, T., {de Bruijne}, J.~H.~J., {et~al.} 2016,
  \aap, 595, A1

\bibitem[{{Gonzalez} {et~al.}(2011){Gonzalez}, {Rejkuba}, {Minniti}, {Zoccali},
  {Valenti}, \& {Saito}}]{gonzalez2011}
{Gonzalez}, O.~A., {Rejkuba}, M., {Minniti}, D., {et~al.} 2011, \aap, 534, L14

\bibitem[{{Gonzalez} {et~al.}(2012){Gonzalez}, {Rejkuba}, {Zoccali}, {Valenti},
  {Minniti}, {Schultheis}, {Tobar}, \& {Chen}}]{gonzalez2012}
{Gonzalez}, O.~A., {Rejkuba}, M., {Zoccali}, M., {et~al.} 2012, \aap, 543, A13

\bibitem[{{Gonzalez} {et~al.}(2015){Gonzalez}, {Zoccali}, {Vasquez}, {Hill},
  {Rejkuba}, {Valenti}, {Rojas-Arriagada}, {Renzini}, {Babusiaux}, {Minniti},
  \& {Brown}}]{gonzalez2015}
{Gonzalez}, O.~A., {Zoccali}, M., {Vasquez}, S., {et~al.} 2015, A\&A, 584, A46

\bibitem[{{Groenewegen} {et~al.}(2008){Groenewegen}, {Udalski}, \&
  {Bono}}]{groenewegen08}
{Groenewegen}, M.~A.~T., {Udalski}, A., \& {Bono}, G. 2008, \aap, 481, 441

\bibitem[{{Harris}(1985)}]{harris85}
{Harris}, H.~C. 1985, \aj, 90, 756

\bibitem[{{Hill} {et~al.}(2011){Hill}, {Lecureur}, {G{\'o}mez}, {Zoccali},
  {Schultheis}, {Babusiaux}, {Royer}, {Barbuy}, {Arenou}, {Minniti}, \&
  {Ortolani}}]{hill2011}
{Hill}, V., {Lecureur}, A., {G{\'o}mez}, A., {et~al.} 2011, \aap, 534, A80

\bibitem[{{Kunder} {et~al.}(2016){Kunder}, {Rich}, {Koch}, {Storm}, {Nataf},
  {De Propris}, {Walker}, {Bono}, {Johnson}, {Shen}, \& {Li}}]{kunder2016}
{Kunder}, A., {Rich}, R.~M., {Koch}, A., {et~al.} 2016, \apjl, 821, L25

\bibitem[{{Luri, Xavier} {et~al.}(2018){Luri, Xavier}, {A Brown, A. G.},
  {Sarro, L.}, {Arenou, F.}, {Bailer-Jones, C. A.L.}, {Castro-Ginard, A.}, {de
  Bruijne, J.}, {Prusti, T.}, {Babusiaux, C.}, \& {Delgado, H.
  E.}}]{gaia_dr2_parallax}
{Luri, Xavier}, {A Brown, A. G.}, {Sarro, L.}, {et~al.} 2018, A\&A

\bibitem[{{Macri} {et~al.}(2015){Macri}, {Ngeow}, {Kanbur}, {Mahzooni}, \&
  {Smitka}}]{macri2015}
{Macri}, L.~M., {Ngeow}, C.-C., {Kanbur}, S.~M., {Mahzooni}, S., \& {Smitka},
  M.~T. 2015, \aj, 149, 117

\bibitem[{{Majaess} {et~al.}(2018){Majaess}, {D{\'e}k{\'a}ny}, {Hajdu},
  {Minniti}, {Turner}, \& {Gieren}}]{majaess2018}
{Majaess}, D., {D{\'e}k{\'a}ny}, I., {Hajdu}, G., {et~al.} 2018, ArXiv
  e-prints, to appear in Ap\&SS [\eprint[arXiv]{1805.04119}]

\bibitem[{{Manick} {et~al.}(2018){Manick}, {Van Winckel}, {Kamath}, {Sekaran},
  \& {Kolenberg}}]{manick2018}
{Manick}, R., {Van Winckel}, H., {Kamath}, D., {Sekaran}, S., \& {Kolenberg},
  K. 2018, ArXiv e-prints [\eprint[arXiv]{1806.08210}]

\bibitem[{{Marconi} \& {Minniti}(2018)}]{marconiminniti2018}
{Marconi}, M. \& {Minniti}, D. 2018, \apjl, 853, L20

\bibitem[{{Matsunaga} {et~al.}(2009){Matsunaga}, {Feast}, \&
  {Menzies}}]{matsunaga2009a}
{Matsunaga}, N., {Feast}, M.~W., \& {Menzies}, J.~W. 2009, \mnras, 397, 933

\bibitem[{{Matsunaga} {et~al.}(2011){Matsunaga}, {Feast}, \&
  {Soszy{\'n}ski}}]{matsunaga11a}
{Matsunaga}, N., {Feast}, M.~W., \& {Soszy{\'n}ski}, I. 2011, \mnras, 413, 223

\bibitem[{{Matsunaga} {et~al.}(2006){Matsunaga}, {Fukushi}, {Nakada},
  {Tanab{\'e}}, {Feast}, {Menzies}, {Ita}, {Nishiyama}, {Baba}, {Naoi},
  {Nakaya}, {Kawadu}, {Ishihara}, \& {Kato}}]{matsunaga06}
{Matsunaga}, N., {Fukushi}, H., {Nakada}, Y., {et~al.} 2006, \mnras, 370, 1979

\bibitem[{{Mauro} {et~al.}(2013){Mauro}, {Moni Bidin}, {Chen{\'e}}, {Geisler},
  {Alonso-Garc{\'{\i}}a}, {Borissova}, \& {Carraro}}]{mauro2013}
{Mauro}, F., {Moni Bidin}, C., {Chen{\'e}}, A.-N., {et~al.} 2013, \rmxaa, 49,
  189

\bibitem[{{McWilliam} \& {Zoccali}(2010)}]{mcwilliamzoccali2010}
{McWilliam}, A. \& {Zoccali}, M. 2010, \apj, 724, 1491

\bibitem[{{Minniti}(1996)}]{minniti1996}
{Minniti}, D. 1996, \apj, 459, 175

\bibitem[{{Minniti} {et~al.}(2010){Minniti}, {Lucas}, {Emerson}, {Saito},
  {Hempel}, {Pietrukowicz}, {Ahumada}, {Alonso}, {Alonso-Garcia}, {Arias},
  {Bandyopadhyay}, {Barb{\'a}}, {Barbuy}, {Bedin}, {Bica}, {Borissova},
  {Bronfman}, {Carraro}, {Catelan}, {Clari{\'a}}, {Cross}, {de Grijs},
  {D{\'e}k{\'a}ny}, {Drew}, {Fari{\~n}a}, {Feinstein}, {Fern{\'a}ndez
  Laj{\'u}s}, {Gamen}, {Geisler}, {Gieren}, {Goldman}, {Gonzalez}, {Gunthardt},
  {Gurovich}, {Hambly}, {Irwin}, {Ivanov}, {Jord{\'a}n}, {Kerins}, {Kinemuchi},
  {Kurtev}, {L{\'o}pez-Corredoira}, {Maccarone}, {Masetti}, {Merlo},
  {Messineo}, {Mirabel}, {Monaco}, {Morelli}, {Padilla}, {Palma}, {Parisi},
  {Pignata}, {Rejkuba}, {Roman-Lopes}, {Sale}, {Schreiber}, {Schr{\"o}der},
  {Smith}, {}, {Soto}, {Tamura}, {Tappert}, {Thompson}, {Toledo}, {Zoccali}, \&
  {Pietrzynski}}]{minniti2010}
{Minniti}, D., {Lucas}, P.~W., {Emerson}, J.~P., {et~al.} 2010, \na, 15, 433

\bibitem[{{Nemec} {et~al.}(1994){Nemec}, {Nemec}, \& {Lutz}}]{nemec1994}
{Nemec}, J.~M., {Nemec}, A.~F.~L., \& {Lutz}, T.~E. 1994, \aj, 108, 222

\bibitem[{{Ness} {et~al.}(2013){Ness}, {Freeman}, {Athanassoula},
  {Wylie-de-Boer}, {Bland-Hawthorn}, {Asplund}, {Lewis}, {Yong}, {Lane}, \&
  {Kiss}}]{ness2013}
{Ness}, M., {Freeman}, K., {Athanassoula}, E., {et~al.} 2013, \mnras, 430, 836

\bibitem[{{Ness} \& {Lang}(2016)}]{nesslang2016}
{Ness}, M. \& {Lang}, D. 2016, \aj, 152, 14

\bibitem[{{Nishiyama} {et~al.}(2009){Nishiyama}, {Tamura}, {Hatano}, {Kato},
  {Tanab{\'e}}, {Sugitani}, \& {Nagata}}]{nishiyama2009}
{Nishiyama}, S., {Tamura}, M., {Hatano}, H., {et~al.} 2009, \apj, 696, 1407

\bibitem[{{Persson} {et~al.}(2004){Persson}, {Madore}, {Krzemi{\'n}ski},
  {Freedman}, {Roth}, \& {Murphy}}]{persson2004}
{Persson}, S.~E., {Madore}, B.~F., {Krzemi{\'n}ski}, W., {et~al.} 2004, \aj,
  128, 2239

\bibitem[{{Pietrukowicz} {et~al.}(2015){Pietrukowicz}, {Koz{\l}owski},
  {Skowron}, {Soszy{\'n}ski}, {Udalski}, {Poleski}, {Wyrzykowski},
  {Szyma{\'n}ski}, {Pietrzy{\'n}ski}, {Ulaczyk}, {Mr{\'o}z}, {Skowron}, \&
  {Kubiak}}]{pietrukowicz2015}
{Pietrukowicz}, P., {Koz{\l}owski}, S., {Skowron}, J., {et~al.} 2015, \apj,
  811, 113

\bibitem[{{Pietrukowicz} {et~al.}(2012){Pietrukowicz}, {Udalski},
  {Soszy{\'n}ski}, {Nataf}, {Wyrzykowski}, {Poleski}, {Koz{\l}owski},
  {Szyma{\'n}ski}, {Kubiak}, {Pietrzy{\'n}ski}, \&
  {Ulaczyk}}]{pietrukowicz2012}
{Pietrukowicz}, P., {Udalski}, A., {Soszy{\'n}ski}, I., {et~al.} 2012, \apj,
  750, 169

\bibitem[{{Pietrzy{\'n}ski} {et~al.}(2013){Pietrzy{\'n}ski}, {Graczyk},
  {Gieren}, {Thompson}, {Pilecki}, {Udalski}, {Soszy{\'n}ski}, {Koz{\l}owski},
  {Konorski}, {Suchomska}, {Bono}, {Moroni}, {Villanova}, {Nardetto},
  {Bresolin}, {Kudritzki}, {Storm}, {Gallenne}, {Smolec}, {Minniti}, {Kubiak},
  {Szyma{\'n}ski}, {Poleski}, {Wyrzykowski}, {Ulaczyk}, {Pietrukowicz},
  {G{\'o}rski}, \& {Karczmarek}}]{pietrzynski2013}
{Pietrzy{\'n}ski}, G., {Graczyk}, D., {Gieren}, W., {et~al.} 2013, \nat, 495,
  76

\bibitem[{{Reid} \& {Brunthaler}(2004)}]{reid2004}
{Reid}, M.~J. \& {Brunthaler}, A. 2004, \apj, 616, 872

\bibitem[{{Ripepi} {et~al.}(2015){Ripepi}, {Moretti}, {Marconi}, {Clementini},
  {Cioni}, {de Grijs}, {Emerson}, {Groenewegen}, {Ivanov}, {Muraveva},
  {Piatti}, \& {Subramanian}}]{ripepi2015}
{Ripepi}, V., {Moretti}, M.~I., {Marconi}, M., {et~al.} 2015, \mnras, 446, 3034

\bibitem[{{Saito} {et~al.}(2012){Saito}, {Hempel}, {Minniti}, {Lucas},
  {Rejkuba}, {Toledo}, {Gonzalez}, {Alonso-Garc{\'{\i}}a}, {Irwin},
  {Gonzalez-Solares}, {Hodgkin}, {Lewis}, {Cross}, {Ivanov}, {Kerins},
  {Emerson}, {Soto}, {Am{\^o}res}, {Gurovich}, {D{\'e}k{\'a}ny}, {Angeloni},
  {Beamin}, {Catelan}, {Padilla}, {Zoccali}, {Pietrukowicz}, {Moni Bidin},
  {Mauro}, {Geisler}, {Folkes}, {Sale}, {Borissova}, {Kurtev}, {Ahumada},
  {Alonso}, {Adamson}, {Arias}, {Bandyopadhyay}, {Barb{\'a}}, {Barbuy},
  {Baume}, {Bedin}, {Bellini}, {Benjamin}, {Bica}, {Bonatto}, {Bronfman},
  {Carraro}, {Chen{\`e}}, {Clari{\'a}}, {Clarke}, {Contreras}, {Corvill{\'o}n},
  {de Grijs}, {Dias}, {Drew}, {Fari{\~n}a}, {Feinstein},
  {Fern{\'a}ndez-Laj{\'u}s}, {Gamen}, {Gieren}, {Goldman},
  {Gonz{\'a}lez-Fern{\'a}ndez}, {Grand}, {Gunthardt}, {Hambly}, {Hanson},
  {He{\l}miniak}, {Hoare}, {Huckvale}, {Jord{\'a}n}, {Kinemuchi}, {Longmore},
  {L{\'o}pez-Corredoira}, {Maccarone}, {Majaess}, {Mart{\'{\i}}n}, {Masetti},
  {Mennickent}, {Mirabel}, {Monaco}, {Morelli}, {Motta}, {Palma}, {Parisi},
  {Parker}, {Pe{\~n}aloza}, {Pietrzy{\'n}ski}, {Pignata}, {Popescu}, {Read},
  {Rojas}, {Roman-Lopes}, {Ruiz}, {Saviane}, {Schreiber}, {Schr{\"o}der},
  {Sharma}, {Smith}, {Sodr{\'e}}, {Stead}, {Stephens}, {Tamura}, {Tappert},
  {Thompson}, {Valenti}, {Vanzi}, {Walton}, {Weidmann}, \&
  {Zijlstra}}]{saito2012}
{Saito}, R.~K., {Hempel}, M., {Minniti}, D., {et~al.} 2012, \aap, 537, A107

\bibitem[{{Saito} {et~al.}(2011){Saito}, {Zoccali}, {McWilliam}, {Minniti},
  {Gonzalez}, \& {Hill}}]{saito2011}
{Saito}, R.~K., {Zoccali}, M., {McWilliam}, A., {et~al.} 2011, \aj, 142, 76

\bibitem[{{Schultheis} {et~al.}(2014){Schultheis}, {Chen}, {Jiang}, {Gonzalez},
  {Enokiya}, {Fukui}, {Torii}, {Rejkuba}, \& {Minniti}}]{schultheis2014}
{Schultheis}, M., {Chen}, B.~Q., {Jiang}, B.~W., {et~al.} 2014, \aap, 566, A120

\bibitem[{{Smith} {et~al.}(2018){Smith}, {Lucas}, {Kurtev}, {Smart}, {Minniti},
  {Borissova}, {Jones}, {Zhang}, {Marocco}, {Contreras Pe{\~n}a}, {Gromadzki},
  {Kuhn}, {Drew}, {Pinfield}, \& {Bedin}}]{smith2018}
{Smith}, L.~C., {Lucas}, P.~W., {Kurtev}, R., {et~al.} 2018, \mnras, 474, 1826

\bibitem[{{Soszy{\'n}ski} {et~al.}(2011){Soszy{\'n}ski}, {Udalski},
  {Pietrukowicz}, {Szyma{\'n}ski}, {Kubiak}, {Pietrzy{\'n}ski}, {Wyrzykowski},
  {Ulaczyk}, {Poleski}, \& {Koz{\l}owski}}]{soszynski2011}
{Soszy{\'n}ski}, I., {Udalski}, A., {Pietrukowicz}, P., {et~al.} 2011, Acta
  Astronomica, 61, 285

\bibitem[{{Soszy{\'n}ski} {et~al.}(2008){Soszy{\'n}ski}, {Udalski},
  {Szyma{\'n}ski}, {Kubiak}, {Pietrzy{\'n}ski}, {Wyrzykowski}, {Szewczyk},
  {Ulaczyk}, \& {Poleski}}]{soszynski08c}
{Soszy{\'n}ski}, I., {Udalski}, A., {Szyma{\'n}ski}, M.~K., {et~al.} 2008, Acta
  Astronomica, 58, 293

\bibitem[{{Soszy{\'n}ski} {et~al.}(2017){Soszy{\'n}ski}, {Udalski},
  {Szyma{\'n}ski}, {Wyrzykowski}, {Ulaczyk}, {Poleski}, {Pietrukowicz},
  {Koz{\l}owski}, {Skowron}, {Skowron}, {Mr{\'o}z}, {Pawlak}, {Rybicki}, \&
  {Jacyszyn-Dobrzeniecka}}]{soszynski2017}
{Soszy{\'n}ski}, I., {Udalski}, A., {Szyma{\'n}ski}, M.~K., {et~al.} 2017, Acta
  Astronomica, 67, 297

\bibitem[{{Spaenhauer} {et~al.}(1992){Spaenhauer}, {Jones}, \&
  {Whitford}}]{spaenhauer1992}
{Spaenhauer}, A., {Jones}, B.~F., \& {Whitford}, A.~E. 1992, \aj, 103, 297

\bibitem[{{Udalski} {et~al.}(2015){Udalski}, {Szyma{\'n}ski}, \&
  {Szyma{\'n}ski}}]{udalski2015}
{Udalski}, A., {Szyma{\'n}ski}, M.~K., \& {Szyma{\'n}ski}, G. 2015, Acta
  Astronomica, 65, 1

\bibitem[{{Valenti} {et~al.}(2016){Valenti}, {Zoccali}, {Gonzalez}, {Minniti},
  {Alonso-Garc{\'{\i}}a}, {Marchetti}, {Hempel}, {Renzini}, \&
  {Rejkuba}}]{valenti2016}
{Valenti}, E., {Zoccali}, M., {Gonzalez}, O.~A., {et~al.} 2016, A\&A, 587, L6

\bibitem[{{Wallerstein}(2002)}]{wallerstein2002}
{Wallerstein}, G. 2002, \pasp, 114, 689

\bibitem[{{Wegg} \& {Gerhard}(2013)}]{wegggerarhd2013}
{Wegg}, C. \& {Gerhard}, O. 2013, \mnras, 435, 1874

\bibitem[{{Willson} \& {Templeton}(2009)}]{willsontempleton2009}
{Willson}, L.~A. \& {Templeton}, M. 2009, in American Institute of Physics
  Conference Series, Vol. 1170, American Institute of Physics Conference
  Series, ed. J.~A. {Guzik} \& P.~A. {Bradley}, 113--121

\bibitem[{{Zoccali} \& {Valenti}(2016)}]{zoccali2016}
{Zoccali}, M. \& {Valenti}, E. 2016, PASA, 33, e025

\bibitem[{{Zoccali} {et~al.}(2017){Zoccali}, {Vasquez}, {Gonzalez}, {Valenti},
  {Rojas-Arriagada}, {Minniti}, {Rejkuba}, {Minniti}, {McWilliam}, {Babusiaux},
  {Hill}, \& {Renzini}}]{zoccali2017}
{Zoccali}, M., {Vasquez}, S., {Gonzalez}, O.~A., {et~al.} 2017, A\&A, 599, A12

\end{thebibliography}

\end{document}